\documentclass[11pt,a4paper]{article} 
\pdfoutput=1
\usepackage{jcappub}
\usepackage{subfigure}
\newcommand{\eq}[1]{\begin{equation}#1\end{equation}}
\newcommand{\eqa}[1]{\begin{eqnarray}#1\end{eqnarray}}
\newcommand{\secs}[1]{\section{#1\label{sec-#1}}}

\newcommand{\fig}[4]{\begin{figure}[#4]\centering\includegraphics[width=#3\textwidth]{Graph-#1.pdf}\caption{#2}\label{fig-#1}\end{figure}}
\newcommand{\figa}[3]{\begin{figure}[#3]\centering #1\caption{#2}\end{figure}}
\newcommand{\figi}[3]{\subfigure[#2]{\includegraphics[width=#3\textwidth]{Graph-#1.pdf}\label{fig-#1}}}

\newcommand{\refeq}[1]{Eq.\ (\ref{eq-#1})}
\newcommand{\refsec}[1]{Section \ref{sec-#1}}

\newcommand{\refig}[1]{Fig.\ \ref{fig-#1}}

\newcommand{\subs}[1]{_\mathrm{#1}}
\newcommand{\sups}[1]{^\mathrm{#1}}
\newcommand{\dd}[1]{d#1}
\newcommand{\mpl}{M\subs{p}}

\newcommand{\cm}[1]{}
\newcommand{\vect}[1]{{\mathbf{#1}}}

\def\ms{M_*}
\def\fNL{f\subs{NL}}
\def\gNL{g\subs{NL}}
\begin{document}
\title{Visible sector inflation and the right thermal history in light of Planck data}

\author{Lingfei Wang,}
\author{Ernestas Pukartas}
\author{and Anupam Mazumdar}
\affiliation{Consortium for Physics, Lancaster University, Lancaster LA1 4YB, UK}

\abstract{Inflation creates perturbations for the large scale structures in the universe, but it also dilutes everything. Therefore it is pertinent that the end of inflation must explain how to excite the Standard Model {\it dof} along with the dark matter. In this paper we will briefly discuss the role of visible sector inflaton candidates which are embedded within the Minimal Supersymmetric Standard Model (MSSM) and discuss their merit on how well they match the current data from the Planck. Since the inflaton carries the Standard Model charges their decay naturally produces all the relevant {\it dof} with no {\it dark/hidden sector radiation} and no isocurvature fluctuations.  We will first discuss a single supersymmetric flat direction model of inflation and demonstrate what parameter space is allowed by the Planck and the LHC.  We will also consider where the perturbations are created by another light field which decays after inflation, known as a {\it curvaton}. The late decay of the curvaton can create observable non-Gaussianity. In the end we will discuss the role of a {\it spectator} field whose origin may not lie within the visible sector physics, but its sheer presence during inflation can still create all the perturbations responsible for the large scale structures including possible non-Gaussianity, while the inflaton is embedded within the visible sector which creates all the relevant matter including dark matter, but no dark radiation.}
\maketitle

\secs{Introduction and motivation for a visible sector inflation}

The primordial inflation~\cite{Guth:1980zm} is the simplest dynamical mechanism which explains the seed perturbations for the cosmic microwave background (CMB) radiation with almost Gaussian perturbations as suggested by the recent Planck data~\cite{Planckc,Planckng,Plancki}. Since inflation dilutes everything other than stretching the initial vacuum fluctuations, after the end of inflation the coherent oscillations of the inflation must excite the Standard Model (SM) quarks and leptons at temperatures sufficiently high to realize SM baryons and dark matter in the current universe~\cite{Mazumdar:2010sa, Allahverdi:2010xz}. In principle, inflation could have occurred in many many phases~\cite{Burgess:2005sb}, and perhaps even future-eternal~\cite{eternal,linde-book}, but it must end in our Hubble patch with the right thermal history and three light neutrino species~\cite{Planckc}.

In principle inflaton whose potential drives inflation could be an arbitrary hidden sector field~\footnote{The sector of particle physics which does not carry the SM charges are typically called the {\it hidden sector}. Any beyond the SM (BSM) physics harbours many hidden sectors which may or may not couple to the SM sector or its minimal extensions. The latter sectors are usually called the {\it visible sector} as they comprise of SM fields or share the SM charges. Inflationary models can be constructed {\it solely} within a visible sector physics, for a review on inflation model buildings, see~\cite{Mazumdar:2010sa}.}, whose properties can be constructed solely to match the observational data  from Planck~\cite{Planckng,Plancki}. However note that the CMB observables {\it merely} probe the gravitational aspect of the problem,  it is not sensitive to the inflation's couplings to the SM matter and neither its origin. Typically hidden sector inflatons are SM gauge singlets~\footnote{There could be more than one inflaton fields and perhaps even of the order of ${\cal O}(10^{2}-10^{3})$ as in the case of {\it assisted inflation}~\cite{Liddle:1998jc}. Although there are some embeddings of such models of inflation within large SU(N) gauge theories~\cite{Jokinen:2004bp}, and in string theory~\cite{strings-A}, but it is highly unlikely that nature would prefer such a route since none of these fields can be embedded within a visible sector physics.}, whose mass and couplings can never be probed directly. In order to explain the universe filled with the SM quarks and leptons such an inflaton should primarily couple {\it only} to the SM sector~\cite{Allahverdi:2007zz}, which is an ad-hoc assumption. A gauge singlet could in principle couple to other sectors, i.e. hidden or visible, there is no symmetry which can completely forbid their couplings to the hidden sector. 

Especially, string theory provides many viable SM gauge singlet inflaton candidates, for a review see~\cite{Linde:2004zz}. Inflation is typically driven either by close string moduli or open string moduli. In either case they are SM gauge singlets - therefore it is not at all clear why and how such an inflaton would decay {\it solely} into the SM {\it dof}. Typically string compactification yields many moduli and hidden sectors~\cite{Polchinski:1998rr}. A high scale inflation, i.e. scale which is higher than the mass of the moduli, could in principle excite all the moduli and dump all its entropy in the hidden sectors fields~\cite{Cicoli:2010ha}. The reason for this is {\it kinematical}, the inflaton can decay into hidden sectors due to typically large branching ratio, i.e. there are more hidden sectors and only one visible SM sector. Furthermore, one might as well worry whether the inflaton could excite dark radiation, provided some of the {\it dof} become extremely light, such as in the case of string axions~\cite{Cicoli:2012aq}, or dark matter, such as in case of Kaluza-Klein dark matter~\cite{Chialva:2012rq}.

In this respect, it is vital that the last phase of primordial inflation, i.e. last $50-60$ e-foldings of inflation must end in a vacuum of BSM physics which can solely excite the relevant SM {\it dof} required for the success of Big Bang Nucleosynthesis (BBN), see for a review~\cite{Pospelov:2010hj}.  In this regard, Minimal Supersymmetric Standard Model (MSSM)~\cite{Chung:2003fi} provides a perfect setup where all the matter content is known and can be probed at the LHC~\cite{ATLAS:2012ae,cms} ~\footnote{In principle the scale of SUSY could be wide ranging, if it is at low scales, such as ${\cal O}(1)~{\rm TeV}$, it can explain the hierarchy problem along with the possibility of directly detecting the SUSY partners of the SM fields at the LHC, for a recent review on naturalness, see~\cite{Feng:2013pwa}.  }. SUSY also helps inflation model building, since  inflation needs a potential which remains sufficiently flat along which the slow-roll inflation can take place in order to generate the observed temperature anisotropy in the CMB. SUSY at any scale guarantees the flatness of such flat directions at a perturbative and a non-perturbative level (for a review see~\cite{Enqvist:2003gh}), besides providing a falsifiable framework at low scales. Furthermore, the lightest SUSY particle can be absolutely stable under R-parity, and thus provides an ideal cold dark matter candidate~\cite{CDM}.

The minimalistic realization would be to embed inflation, dark matter within MSSM which are all determined by the {\it known} SM couplings which provides credibility not only to particle physics but also to cosmological predictions. Our aim of this paper will be to show this within three paradigms for the inflationary cosmology -- in all the cases inflation happens in the visible sector of MSSM. The origin of perturbations could be sourced from the visible sector physics or it might as well arise from the hidden sector, we will discuss the role of hidden sector here which might be responsible for creating mild non-Gaussianity. We will also discuss their merits in conjunction with the release of the Planck data along with the constraints arising from the LHC.

\secs{Three paradigms within visible sector inflation}
One can envisage three realistic scenarios. A simple single field model of inflation and a scenario with multi-fields.  In the latter case  we can capture all the essence by mimicking just two fields -- one which is inflaton and the other could be either curvaton~\cite{david,enqvist,moroi}, or a spectator field~\cite{Mazumdar:2012rs,Wang:2013oea} as the simplest examples.  

\begin{itemize}

\item A single field model of inflation:\\
It is well known that a single field model of inflation with a canonical kinetic term will yield almost Gaussian perturbations. Of course, one can depart from the simplest assumptions to generate non-Gaissianity, such as sudden change in the potential, modifying the initial vacuum from Bunch-Davis, or introducing non-canonical kinetic term, for a review on non-Gaussianity see~\cite{Bartolo}. All these have interesting consequences for the primordial non-Gaussianity, but most of them are severely constrained by the current observations~\cite{Planckng}~\footnote{It is possible to obtain small bispectrum but large trispectrum in the CMB in the cyclic universe scenario~\cite{Biswas:2013fna}.}.

In this paper we will revisit the parameter space of  a visible sector single field model of inflation embedded within MSSM with canonical kinetic term and with the Bunch-Davis initial vacuum condition~\cite{Allahverdi:2006iq,Allahverdi:2006we,Allahverdi:2006cx}. In all these models inflation happens below the Planck scale and generate small non-Gaussianities. Furthermore, these models produce the right thermal history of the universe without any dark radiation.

\item Curvaton scenario:\\
A light subdominant field during inflation can also seed the perturbations for the CMB. In the simplest scenarios it is assumed that the inflaton fluctuations are sub-dominant. The light field known as a curvaton~\cite{david,enqvist,moroi} can slow roll after the end of inflation, and decays later on once the inflaton has completely decayed. While decaying the curvaton converts its initial isocurvature fluctuations into curvature perturbations. This conversion leads to a pure adiabatic fluctuations if the curvaton dominates while decaying, on the other hand if the curvaton decay products are sub-dominant compared to the energy density of the inflaton decay products, then there is a residual isocurvature fluctuations. Furthermore since the conversion itself is non-adiabatic, there is a generation of non-Gaussian perturbations of the local configuration. In order not to generate residual isocurvature fluctuations, the inflaton decay products must thermalize with that of the curvaton decay products. A priori this is a non-trivial condition. The only way it could be satisfied provided  both inflaton and curvaton can be embedded within the visible sector, i.e. MSSM,  then this problem could be addressed amicably since both the fields would decay into the MSSM {\it dof}~\cite{Mazumdar:2011xe}. 

\item Spectator scenario:\\
This is a completely new paradigm where a light subdominant field like curvaton is present during inflation, but it decays into radiation much before the end of inflation~\cite{Mazumdar:2012rs,Wang:2013oea}. The sheer presence of such a light field can create perturbations for the CMB, but since the field decays during inflation, its decay products need not be that of the SM or MSSM {\it dof}. In principle if inflation is occurring within a visible sector the perturbations can be seeded by the hidden sector field, which is advantageous for many theories of BSM including string theory. We will illustrate this for the first time with an example of inflation occurring within MSSM, while the spectator field is made up of arbitrary gauge singlet arising from the hidden sector physics.
\end{itemize}

There could be two possibilities for the observed tensor to scalar ratio being negligible ($r<0.11$ with 95\% CL)~\cite{Plancki}. The scale of inflation could be genuinely below the GUT scale, which is the case we will be considering in all the examples below, or the second option could be  that the {\it gravity} is purely classical and so is the vacuum~\cite{Ashoorioon:2012kh}, while matter component is treated quantum mechanically; for a review on cosmological perturbation, see~\cite{Mukhanov:1990me}. A linearized Einstein gravity has no source term, therefore for a classical gravity {\it without} any source for exciting gravity waves in a homogeneous and isotropic universe, the resultant primordial gravitational waves will be absolutely zero~\cite{Ashoorioon:2012kh}. Any positive detection of primordial gravitational waves will indeed shed an important light on whether gravity should be treated classically or not.

\section{Inflection point potential for MSSM flat directions}
The MSSM provides nearly 300 gauge-invariant $F$-and $D$-flat directions~\cite{Gherghetta:1995dv,Dine:1995kz}, which are all charged under the SM gauge group. Out of these flat directions, there are particularly 2 $D$-flat directions: $\widetilde{u}\widetilde{d}\widetilde{d}$ and $\widetilde{L}\widetilde{L}\widetilde{e}$, which carry the SM charges and  can be the ideal inflaton candidates~\cite{Allahverdi:2006iq,Allahverdi:2006we,Allahverdi:2006cx}, where $\widetilde u,~\widetilde d$ correspond to the right handed squarks, $\widetilde L$ corresponds to the left handed slepton, and $\widetilde e$ corresponds to the right handed selectron.  Both the inflaton candidates provide {\it inflection point} in their respective potentials where inflation can be driven for sufficiently large e-foldings of inflation to explain the current universe and explain the seed perturbations for the temperature anisotropy in the CMB~\cite{Allahverdi:2006iq,Allahverdi:2006we,Allahverdi:2006cx}, see also \cite{Martin:2013tda}.

Since both $\widetilde{u}\widetilde{d}\widetilde{d}$ and $\widetilde{L}\widetilde{L}\widetilde{e}$ flat directions are lifted by higher order superpotential terms of the following form, which would provide non-vanishing $A$-term in the potential even at large VEVs, but below the cut-off scale: 
\begin{equation} \label{eq-i-supot}
W \supset {\lambda \over 6}{\Phi^6 \over \mpl^3}\, ,
\end{equation}
where $\lambda \sim {\cal O}(1)$~\footnote{~The exact value of $\lambda$ is irrelevant for the CMB analysis, as it does not modify the CMB predictions. However it is possible to extract $\lambda$ by integrating out the heavy {\it dof}. In the case if the origin of these operators arise from either $SU(5)$ or $SO(10)$, then the typical value is of order $\lambda \sim {\cal O}(10^{-2})$ for $SO(10)$ and $\lambda \sim {\cal O}(1)$ for $SU(5)$, as shown in Ref.~\cite{Allahverdi:2007vy}.}, and $\mpl=2.4\times 10^{18}\,\mathrm{GeV}$ is the reduced Planck mass. The scalar component of $\Phi$ superfield, denoted by $\phi$, is given by~\footnote{The representations for the flat directions are given by:
$\widetilde u^{\alpha}_i=\frac1{\sqrt{3}}\phi\,,~
\widetilde d^{\beta}_j=\frac1{\sqrt{3}}\phi\,,~
\widetilde d^{\gamma}_k=\frac{1}{\sqrt{3}}\phi\,$
Here $1 \leq \alpha,\beta,\gamma \leq 3$ are color indices, and $1
\leq i,j,k \leq 3$ denote the quark families. The flatness constraints
require that $\alpha \neq \beta \neq \gamma$ and $j \neq k$.
$\widetilde L^a_i=\frac1{\sqrt{3}}\left(\begin{array}{l}0\\ \phi\end{array}\right)\,,~
\widetilde L^b_j=\frac1{\sqrt{3}}\left(\begin{array}{l}\phi\\ 0\end{array}\right)\,,~
\widetilde e_k=\frac{1}{\sqrt{3}}\phi\,,$
where $1 \leq a,b \leq 2$ are the weak isospin indices and $1 \leq
i,j,k \leq 3$ denote the lepton families. The flatness constraints
require that $a \neq b$ and $i \neq j \neq k$. Note that the cosmological perturbations do not care which combination arises, as gravity couples universally.
}
\begin{equation} \label{eq-i-infl}
\phi = {\widetilde{u} + \widetilde{d} + \widetilde{d} \over \sqrt{3}} \,, ~~~~~~ ~ ~ \phi = {\widetilde{L} + \widetilde{L} + 
\widetilde{e} \over \sqrt{3}},
\end{equation}
for the $\widetilde{u}\widetilde{d}\widetilde{d}$ and $\widetilde{L}\widetilde{L}\widetilde{e}$ flat directions respectively.
After minimizing the potential along the angular direction $\theta$ ($\Phi$ = $\phi e^{i \theta}$), we can situate the real part of $\phi$ by rotating it to the corresponding angles $\theta_{\rm min}$. The scalar potential is then found to be~\cite{Allahverdi:2006iq,Allahverdi:2006we}
\begin{equation} \label{eq-i-scpot}
V(\phi) = {1\over2} m^2_\phi\, \phi^2 - A {\lambda\phi^6 \over 6\,\mpl^{3}} + \lambda^2
{{\phi}^{10} \over \mpl^{6}}\,,
\end{equation}
where $m_\phi$ and $A$ are the soft breaking mass and the $A$-term respectively ($A$ is a positive quantity since its phase is absorbed by a redefinition of $\theta$ during the process). The masses for $\widetilde{L}\widetilde{L}\widetilde{e}$ and $\widetilde{u}\widetilde{d}\widetilde{d}$ are given by:
\begin{eqnarray}\label{eq-i-masses}
m^2_{\phi}=\frac{m^2_{\widetilde L}+m^2_{\widetilde L}+m^2_{\widetilde e}}{3}\,,\\
m^2_{\phi}=\frac{m^2_{\widetilde u}+m^2_{\widetilde d}+m^2_{\widetilde d}}{3}\,.
\end{eqnarray}
Note that the masses are now VEV dependent, i.e. $m^2(\phi)$. The inflationary perturbations will be able to constrain 
the inflaton mass only at the scale of inflation, i.e. $\phi_0$, while LHC will be able to constrain the masses at the LHC scale. However both the physical 
quantities are related to each other via RGE as we will discuss below.
For
\begin{equation} \label{eq-i-dev}
{A^2 \over 40 m^2_{\phi}} \equiv 1 - 4 \alpha^2\, ,
\end{equation}
where $\alpha^2 \ll 1$, there exists a point of inflection ($\phi_0$) in $V(\phi)$, where
\begin{eqnarray}
&&\phi_0 = \left({m_\phi \mpl^{3}\over \lambda \sqrt{10}}\right)^{1/4} + {\cal O}(\alpha^2) \, , \label{eq-i-infvev} \\
&&\, \nonumber \\
&&V^{\prime \prime}(\phi_0) = 0 \, , \label{eq-i-2nd}
\end{eqnarray}
at which
\begin{eqnarray}
\label{eq-i-pot}
&&V(\phi_0) = \frac{4}{15}m_{\phi}^2\phi_0^2 + {\cal O}(\alpha^2) \, , \\
\label{eq-i-1st}
&&V'(\phi_0) = 4 \alpha^2 m^2_{\phi} \phi_0 \, + {\cal O}(\alpha^4) \, , \\
\label{eq-i-3rd}
&&V^{\prime \prime \prime}(\phi_0) = 32\frac{m_{\phi}^2}{\phi_0} + {\cal O}(\alpha^2) \, .
\end{eqnarray}
From now on we only keep the leading order terms in all expressions. 
Note that inflation occurs within an interval~\footnote{For a low scale inflation, setting the initial condition is always challenging. However in the case of a MSSM or string theory landscape where there are many false vacua at high and high scales, then it is conceivable that earlier phases of inflation could have occurred in those false vacua. This large vacuum energy could lift the fMSSM lat direction condensate either via quantum fluctuations~\cite{Allahverdi:2007wh}, or via classical initial condition which happens at the level of background without any problem, see~\cite{Allahverdi:2008bt}.}
\begin{equation} \label{eq-i-plateau}
\vert \phi - \phi_0 \vert \sim {\phi^3_0 \over 60 \mpl^2} ,
\end{equation}
in the vicinity of the point of inflection, within which the slow roll parameters $\epsilon \equiv (\mpl^2/2)(V^{\prime}/V)^2$ and $\eta \equiv \mpl^2(V^{\prime \prime}/V)$  are smaller than $1$. The Hubble expansion rate during inflation is given by
\begin{equation}\label{eq-i-hubble}
H_{inf} \simeq \frac{2}{\sqrt{45}}\frac{m_{\phi}\phi_0}{M_{p}}\,.
\end{equation}
In order to obtain the flat potential, it is crucial that the $A(\phi_0)$-term ought to be close to $m_{\phi}(\phi_0)$ in the above potential \refeq{i-scpot}. This can be obtained within two particular scenarios -- (1) Gravity Mediation: 
in gravity-mediated SUSY breaking, the $A$-term and the soft SUSY breaking mass are of the same order of magnitude as the gravitino mass, i.e. $m_{\phi} \sim A \sim m_{3/2} $~\cite{Nilles:1983ge}, and (2) Spilt SUSY: in Split SUSY scenario the scale of SUSY is high and sfermions are very heavy, the $A$-term is typically protected by R-symmetry, see Refs.~\cite{ArkaniHamed:2004fb,Giudice:2004tc}, as a result the $A$-term could be very small compared to the soft masses. However, if the Yukawa hierarchy arises from the Froggatt-Nielsen mechanism, then the $A$-term can be made as large as that of the soft mass, i.e. $m_\phi \sim A$, as in the case of Ref.~\cite{Babu:2005ui}.


\begin{figure}[t]
\includegraphics[width=\textwidth]{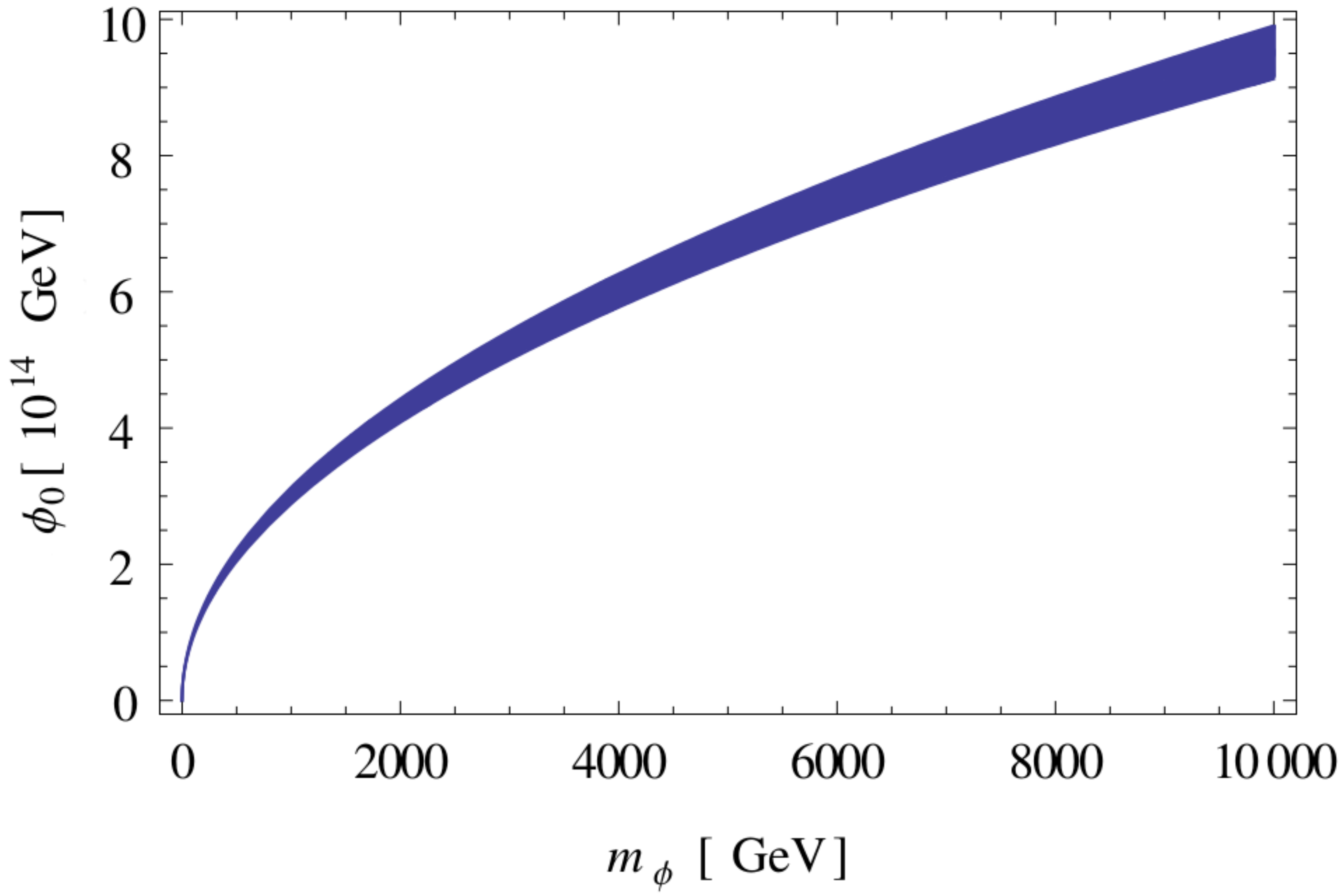}
\caption{$(\phi_0,m_{\phi})$ plane in which inflation is in agreement with the cosmological observations of the temperature anisotropy of the CMB fluctuations. The blue region shows the inflaton energy scale and inflaton mass which are compatible with the central value of the amplitude of the seed perturbations, $P_{\zeta}=2.196\times10^{-9}$, and the $1\sigma$ allowed range of spectral tilt $n_s=0.9603\pm0.0073$~\cite{Planckc}.}
\label{fig-phi-mphi-infl}
\end{figure}


The above potential \refeq{i-scpot} has been studied extensively in Refs.~\cite{Allahverdi:2006we,Bueno Sanchez:2006xk,Enqvist:2010vd}. The amplitude of density perturbations $\delta_H$ and the scalar spectral index $n_s$ are given by:
\begin{equation} \label{eq-i-ampl}
\delta_H =\frac{2}{5}\sqrt{P_\zeta}= {8 \over \sqrt{5} \pi} {m_{\phi} M_{p} \over \phi^2_0}{1 \over \Delta^2}
~ {\rm sin}^2 [{\cal N}_{\rm COBE}\sqrt{\Delta^2}]\,, \end{equation}
and
\begin{equation} \label{eq-i-tilt}
n_s = 1 - 4 \sqrt{\Delta^2} ~ {\rm cot} [{\cal N}_{\rm COBE}\sqrt{\Delta^2}], \end{equation}
respectively, where
\begin{equation} \label{eq-i-Delta}
\Delta^2 \equiv 900 \alpha^2 {\cal
N}^{-2}_{\rm COBE} \Big({M_{p} \over \phi_0}\Big)^4\,. \end{equation}
In the above, ${\cal N}_{\rm COBE}$ is the number of e-foldings between the time when the observationally relevant perturbations are generated till the end of inflation and follows: 
${\cal N}_{\rm COBE} \simeq 66.9 + (1/4) {\rm ln}({V(\phi_0)/\mpl^4}) \sim 50$. 
The running of the spectral tilt is negligible~\cite{Allahverdi:2006we,Bueno Sanchez:2006xk,Enqvist:2010vd} within the current bound of the Planck observations~\cite{Plancki}. The perturbations are due to single canonical field, therefore one would not expect large non-Gaussianity from this model. The observed non-Gaussianity parameter denoted by $\fNL\leq 1$ is bounded by the slow roll parameters, see Ref.~\cite{Bartolo}, and is consistent with Planck~\cite{Planckng}. The scale of inflation is low enough that one would not expect any observed tensor perturbations in any future CMB experiments~\footnote{In this paper we will mostly consider this scenario. In order to obtain large observable tensor to scalar ratio, $r$, we will have to embed this model within $N=1$ supergravity (SUGRA). This will modify the potential with a large vacuum energy density besides providing SUGRA corrections to mass and A-term~\cite{Mazumdar:2011ih,Hotchkiss:2011gz}. It has been shown that it is possible to obtain $r\sim 0.05$ for both inflaton flat directions: $\widetilde u\widetilde d\widetilde d$ and $\widetilde L\widetilde L\widetilde e$~\cite{Hotchkiss:2011gz}.}.

Instant reheating and thermalization~\cite{Felder} occurs when a single MSSM flat direction is responsible for inflation. 
This is due to the gauge couplings of the inflaton to gauge/gaugino fields. Within $10-20$ inflaton oscillations radiation-dominated universe prevails, as shown in Ref.~\cite{Allahverdi:2011aj}. The resultant reheat temperature at which all the MSSM {\it dof} are in thermal equilibrium (kinetic and chemical equilibrium) is given by~\cite{Allahverdi:2011aj}
\begin{equation}
T_{rh}\sim 2\times 10^{8}~{\rm GeV}.
\end{equation}
Since the temperature of the universe is so high, it immediately thermalizes the LSP provided it has gauge interactions. The LSP relic density is then given by the Standard (thermal) Freeze-out mechanism.
In particular, if the neutralino is the LSP, its relic density is determined by its annihilation and coannihilation rates~\cite{Allahverdi:2007vy,Boehm:2012rh}. 

The advantage of realizing inflation in the visible sector is that it is possible to {\it nail} down the thermal history of the universe precisely. 
At temperatures below $10-100$~GeV there will be no extra degrees of freedom in the thermal bath except that of the SM, therefore BBN can proceed without any trouble within low scale SUSY scenario. This reheat temperature is marginally compatible with the BBN bound for the gravitino mass $m_{3/2}\geq {\cal O}(\rm TeV)$. It is also sufficiently high that various mechanisms of baryogenesis may be invoked to generate the observed baryon asymmetry of the universe.

In \refig{phi-mphi-infl} we have explored a wide range of the inflaton mass, $m_{\phi}$, where inflation can explain the observed 
temperature anisotropy in the CMB with the right amplitude,
 $P_\zeta=2.196\times10^{-9}$, and the tilt in the power spectrum, $n_s=0.9603\pm0.0073$~\cite{Planckc}.  The observables 
 $P_\zeta$ and $n_s$ have been shown by blue region. We have restricted ourselves to VEV below the GUT scale. Within the current parameter range the model provides negligible running in the tilt which is well within the observed limit.  We have allowed a wide range for $m_{\phi}$ and $\phi_0$ (the inflection point) to show that  inflation can indeed happen within SUSY from low scales to high scale SUSY breaking soft-masses. High scale soft masses could be made compatible within split-SUSY scenario~\cite{Babu:2005ui}.

Using renormalization group equations the mass of the inflaton can be evaluated at any energy scales, thus 
providing connection between physics at the very high energies in early universe and experimentally probed scales at LHC. 
For the $\widetilde u\widetilde d\widetilde d$ flat direction RGE is~\cite{Allahverdi:2007vy,Boehm:2012rh}:
\begin{equation}
\begin{aligned}
\label{eq-i-rgudd}
&\hat{\mu} \frac{dm^2_\phi}{d\hat{\mu}}=-\frac{1}{6\pi^2}\bigg(4M_3^2g_3^2+\frac{2}{5}M_1^2g_1^2\bigg),
\\&\hat{\mu} \frac{dA}{d\hat{\mu}}=-\frac{1}{4\pi^2}\bigg(\frac{16}{3}M_3g_3^2+\frac{8}{5}M_1g_1^2\bigg),
\end{aligned}
\end{equation}
where  $\hat{\mu} = \hat\mu_0=\phi_0$ is the VEV at which inflation 
occurs. For $\widetilde{L}\widetilde{L}\widetilde{e}$:
\begin{equation}
\begin{aligned}
\label{eq-i-rglle}
&\hat\mu\frac{dm^2_\phi}{d\hat\mu}=-\frac{1}{6\pi^2}\bigg(\frac{3}{2}M_2^2g_2^2+\frac{9}{10}M_1^2g_1^2\bigg),
\\&\hat\mu\frac{dA}{d\hat\mu}=-\frac{1}{4\pi^2}\bigg(\frac{3}{2}M_2g_2^2+\frac{9}{5}M_1g_1^2\bigg),
\end{aligned}
\end{equation}
where $M_1$, $M_2$, $M_3$ are $U(1)$, $SU(2)$ and $SU(3)$ gaugino masses, which all, assuming SUSY models which obey universality conditions like constrained MSSM (CMSSM)~\cite{mSUGRA}, 
equate to $m_{1/2}$ at the unification scale, and $g_1$, $g_2$ and $g_3$ are the associated couplings. To solve these equations, one needs to take into 
account of the running of the gaugino masses and coupling constants which are given by, see \cite{Nilles:1983ge}:
\begin{equation}
\beta (g_i)=\alpha_i g_i^3 \hspace{1.5cm}
\beta\bigg{(}\frac{M_i}{g_i^2}\bigg{)}=0,
\end{equation}
with $\alpha_1={11}/{16\pi^2}$, $\alpha_2={1}/{16\pi^2}$ and $\alpha_1=-{3}/{16\pi^2}$. 

\begin{figure}[t]
\includegraphics[width=1.0\linewidth]{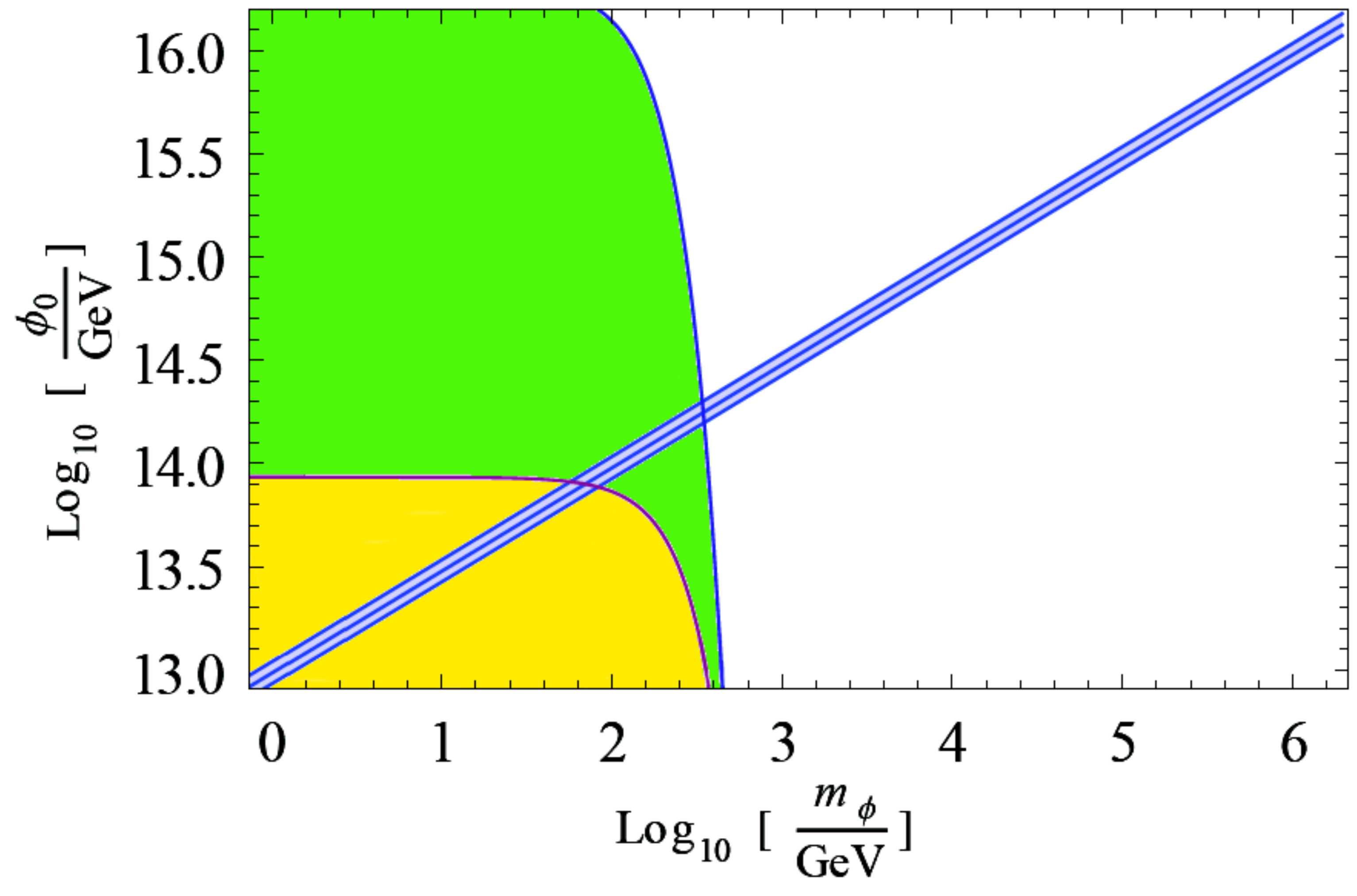}
\caption{$(\phi_0,m_{\phi})$ plane showing the inflationary parameter space that may be ruled out in a likely case if SUSY is not found below $1\,\mathrm{TeV}$. Green region denotes the exclusion if the inflaton is $\widetilde{u}\widetilde{d}\widetilde{d}$ and yellow is for the $\widetilde{L}\widetilde{L}\widetilde{e}$ case. The blue band shows the $(\phi_0,m_\phi)$ values which are compatible with the central value of the amplitude of the seed perturbations, $P_{\zeta}=2.196\times10^{-9}$, and the $3\sigma$ allowed range of spectral tilt $n_s=0.9603\pm0.0219$~\cite{Planckc}.}
\label{fig-phi-mphi}
\end{figure}

Within CMSSM one can try to constrain the inflaton mass for ${\widetilde u}{\widetilde d}{\widetilde d}$ and $\widetilde L\widetilde L\widetilde e$ similar to the analysis of Ref.~\cite{Boehm:2012rh}. The current LHC searches for SUSY particles put a stringent limit on  squarks and sleptons, see~\cite{SUSY-searches1,SUSY-searches2}, and as a result on the inflaton mass as shown in \refig{phi-mphi}.~\footnote{A very similar analysis could be carried out for the MSSM Higgs inflation: $\mu H_uH_d$~\cite{Chatterjee:2011qr}.}


\section{Inflation and curvaton both embedded within MSSM\label{sec-curvaton}}

The curvaton scenario~\cite{david,enqvist,moroi} is an alternative mechanism for creating perturbations. In this scenario, the density perturbations are sourced by the quantum fluctuations of a light scalar field $\sigma$, known as the curvaton, which makes a negligible contribution to the energy density during inflation and decays after the decay of the inflaton field $\phi$. The advantage of the curvaton mechanism is that it can generate non-Gaussianity \cite{david,nG-curvaton} in the primordial density perturbations and also significant residual isocurvature perturbations, neither of which are possible in the usual single-field inflation models. Both signatures are now well constrained by the current Planck data~\cite{Planckc,Planckng}.

If the curvaton does not completely dominate the energy density at the time of its decay, the process of conversion of initial isocurvature perturbations into adiabatic curvature perturbations can enhance the local form of non-Gaussian fluctuations by 
\eq{f_\mathrm{NL}\sim \frac{1}{r}\,,~~ {\rm for}~~ r< 1\,,}
where
\eq{r\equiv\frac{\rho_{\sigma}}{\rho_{\sigma}+\rho_\gamma}}
is the curvaton's energy density ratio at the time the curvaton decays~\cite{david}. Here $\rho_\gamma$ is the energy density of the radiation as the decay products of the inflaton.

However, if either the curvaton or the inflaton belongs to a hidden sector of BSM, they may decay into other fields beyond the SM {\it dof}. There is no guarantee that the hidden and visible sector {\it dof}  would reach thermal equilibrium before the BBN takes place. In this case, residual anti-correlated isocurvature perturbations are expected to be in conflict with the CMB data, which constrain them to be $\lesssim5\%$~\cite{Plancki}. If the curvaton belongs to the visible sector but the inflaton does not, a value of $r\sim1$ would avoid this conflict \cite{Curvaton-enqvist}, but would render any non-Gaussianity undetectable. Note that if $r\sim1$ the curvaton is solely responsible for exciting all the SM {\it dof}, so it must carry the SM charges in order to avoid {\it dark radiation} for instance~\cite{Curvaton-enqvist,Kasuya:2004}. The curvaton scenario lends strong support to a visible sector dark matter such as neutralino in the case of the LSP, because either from the decay of the inflaton or from the curvaton, the neutralino would thermalize with the rest of the plasma soon after its decay, and its final abundance  will be determined by its annihilation and co-annihilation rates.

Keeping all these constraints in mind we need to embed both inflaton and curvaton within a visible sector of BSM physics where they both decay into the SM {\it dof}. Let us consider the case where the inflaton, $\phi$, and the curvaton, $\sigma$, both originate from different saddle point directions which are orthogonal to each other at least at the lowest orders in an effective field theory~\footnote{In Ref.~\cite{Mazumdar:2011xe} we embedded inflation and curvaton both within visible sector, i.e. within MSSM, for the first time. We found that if the scale of inflation was higher than the effective mass of the curvaton then the spectral tilt would tend towards flat spectrum, i.e.  $n_s\sim 1$. In this section we will illustrate with the same potential how to obtain a spectral tilt close to $n_s=0.9603$ and the required local non-Gaussianity in the range of the Planck data~\cite{Plancki,Planckng}. }. The total potential is
\eq{V_{tot}\equiv V(\phi)+U(\sigma).}

Let us first discuss the origin of the curvaton, which we take to be an $R$-parity conserving $D$-flat direction of the MSSM. For the purpose of illustration we consider that to be ${\widetilde L}{\widetilde L}{\widetilde e}$, which is lifted by the non-renormalizable operator:
\eq{W\supset \frac{\lambda}{6}\frac{\Sigma^{6}}{M_{\ast}^{6}}\,,\label{eq-c-supot1}}
where $\lambda$ is a non-renormalizable coupling induced by integrating out the heavy fields at the intermediate scale, $M_\ast$, which could be close to the GUT scale, i.e. $M_\ast \sim M_{GUT}$. The scalar component of the $\Sigma$ superfield and its mass are given
by:
\begin{equation}
\sigma=\frac{({\widetilde L}+{\widetilde L}+{\widetilde e})}{\sqrt{3}}\,,~~~~~~~~~~m^2_\sigma=
\frac{m^2_{\widetilde L}+m^2_{\widetilde L}+m^2_{\widetilde e}}{3}\,.
\end{equation}
where at the lowest order the potential along the $\sigma$ direction is given by similar to 
Eq.~(\ref{eq-i-scpot})~\footnote{Note that this curvaton potential has the same origin that of MSSM inflation discussed in \refeq{i-scpot}.}:
\begin{equation}\label{eq-c-scpot-11}
U(\sigma)=\frac{1}{2}m_{\sigma}^{2}|\sigma|^{2}-\frac{A\lambda}{6}\frac{\sigma^{6}}{M_{\ast}^{3}}+\lambda^{2}\frac{|\sigma|^{10}}{M_{\ast}^{6}}\,,
\end{equation}
where $A\sim m_{\sigma}\sim \mathcal{O}(1-10)$~TeV, are the soft SUSY-breaking terms.
We will assume that the curvaton rolls on a saddle point of the potential, i.e. $A=\sqrt{40}m_\sigma$, so the saddle point lies at
\eq{\sigma_0=\left(\frac{m_\sigma}{\sqrt{10}\lambda\ms}\right)^{1/4}\ms.}

We now turn to the origin of $V(\phi)$ within the MSSM. Let us consider a flat-direction orthogonal to the curvaton. 
If the curvaton is ${\widetilde L}{\widetilde L}{\widetilde e}$, the inflaton could be ${\widetilde u}{\widetilde d}{\widetilde d}$ direction. 
In which case {\it both} inflaton and curvaton are embedded within MSSM. 
We take the inflaton direction to be squarks, typically they are expected to be heavier than the sleptons: 
\begin{equation}
\phi=\frac{{\widetilde u}+{\widetilde d}+{\widetilde d}}{\sqrt{3}}\,.
\end{equation}
 Note that $\widetilde u\widetilde d\widetilde d$ and $\widetilde L\widetilde L\widetilde e$ remain two  {\it independent} directions for the entire range of VEVs. This flat direction will also be lifted by the non-renormalizable operators. However, at larger VEVs the potential energy density stored in the ${\widetilde u\widetilde d\widetilde d}$ direction will be larger than that of  ${\widetilde L\widetilde L\widetilde e}$, so it would be lifted by higher order terms: 
\begin{equation}
\label{eq-c-supot2}
W =\sum_{m\geq2}\frac{\lambda_{m}}{3m}\frac{\Phi^{3m}}{M_{\ast}^{3m-3}}\,.
\end{equation}
The potential at lowest order would be:
\begin{equation}
\label{eq-c-catpot}
V(\phi) =\left |\lambda_{2}\frac{\phi^{5}}{M_{\ast}^{3}}+\lambda_{3}\frac{\phi^{8}}{M_{\ast}^{6}}+
\lambda_{4}\frac{\phi^{11}}{M_{\ast}^{9}}+\cdots \right |^{2}
\end{equation}
where $\cdots$ contains the higher order terms. Note that the $\lambda_m$ in  \refeq{c-supot2} are all non-renormalizable couplings induced  by integrating out the heavy fields at the intermediate scale. At energies below the cut-off scale these coefficients need not necessarily be of $\mathcal{O}(1)$. 

Potentials like \refeq{c-catpot} were studied in Refs.~\cite{Dutta,Dutta-1}. For $\lambda_{2}\ll \lambda_{3}\ll \lambda_{4}\ll \lambda_{n}\leq \mathcal{O}(1)$, they provide a unique solution for which the first and second order derivatives of the potential vanish along both radial and angular direction in the complex plane: $\partial V/\partial\phi=\partial V/\partial \phi^{\ast}=\partial^{2}V/\partial\phi^{2}=\partial^{2}V/\partial\phi^{\ast 2}=0$ (a saddle point condition). For the first three terms in \refeq{c-catpot}, it is possible to show that this happens when 
\begin{equation}\label{eq-c-cond}
\lambda_{3}^{2}=\frac{55}{16} \lambda_{2} \lambda_{4}\,,
\end{equation}
at the VEVs: $\phi=\phi_{0} \exp{[i\pi/3,~i\pi,~i5\pi/3]}$, where
\eq{\phi_{0}=\left(\frac{5\lambda_2}{11\lambda_4}\right)^\frac{1}{6}M_{\ast}.}
Concentrating on the real direction, the potential energy density stored in the inflaton sector is given by:
\begin{equation}
\label{eq-c-relation}
 V(\phi_0)\sim \left(\frac{9}{44}\right)^{2}\lambda_{2}^{2}\frac{\phi_{0}^{10}}{M_{\ast}^{6}}\,,
 \end{equation}
where $\phi_{0}\ll M_\ast$.


\fig{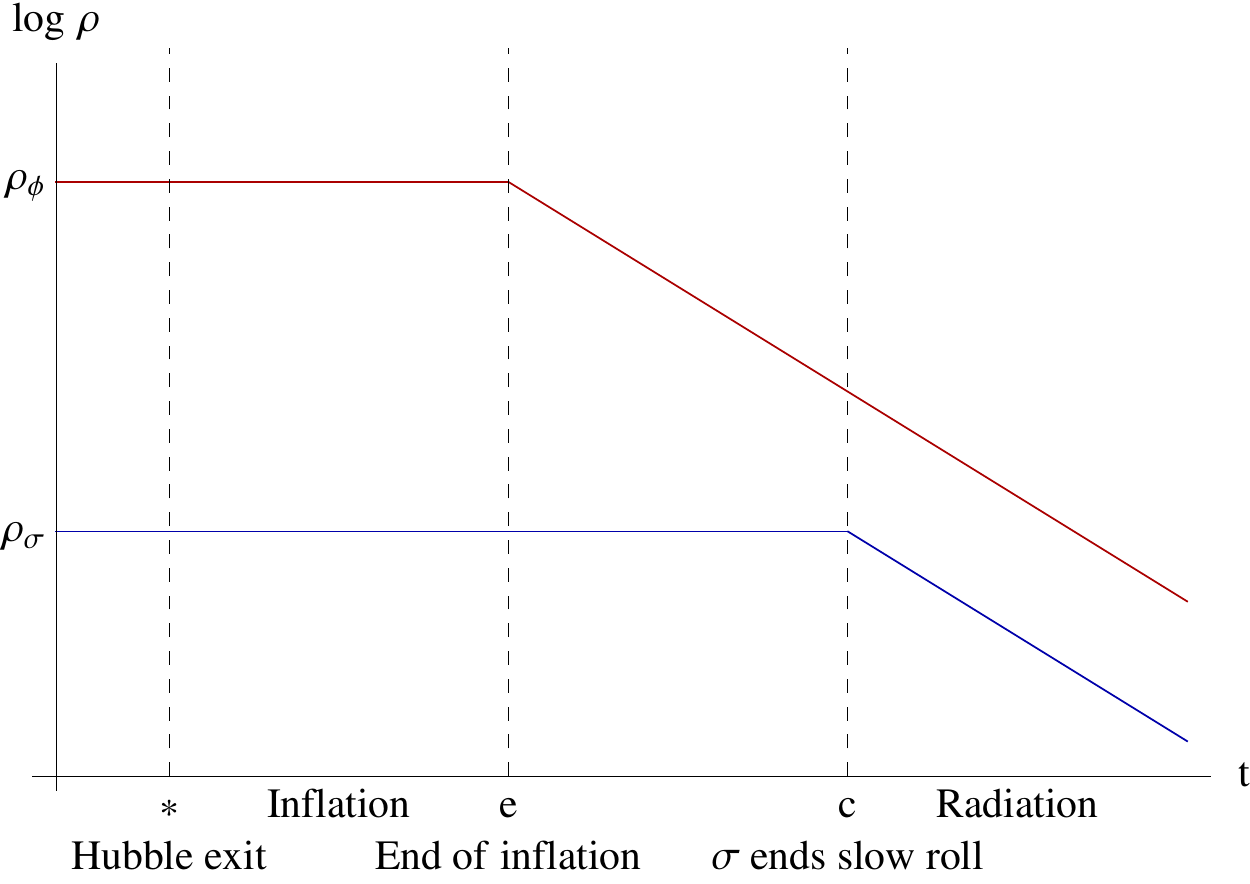}{A schematic timeline is shown for the curvaton scenario. The red and blue curves correspond to the energy densities of the inflaton $\phi$ and curvaton $\sigma$ (or their decay products) respectively. In both cases the flat directions decay into MSSM relativistic \emph{dof} in less than one Hubble time.}{0.7}{}


With the inflaton and curvaton potentials, the universe would then undergo the evolution as indicated in \refig{C-Timeline}. The curvaton $\sigma$ only ends slow roll some time after inflation. In our case both $\phi$ and $\sigma$ will decay within one Hubble time of the evolution. Both carry the SM gauge charges for which thermalization is similar to the case of instant preheating as shown in Refs.~\cite{Allahverdi:2011aj, Felder}. As is also shown in \refig{C-Timeline}, we will use subscripts ``$*$'', ``$e$'', and ``$c$'' to indicate the Hubble exit of the relevant perturbations, the end of inflation, and the time when $\sigma$ decays, respectively.

During inflation, both fields are slowly rolling. The curvaton $\sigma$ remains subdominant so inflation is totally determined by the inflaton $\phi$, where the e-folds from the Hubble exit of relevant modes to the end of inflation is $N_1$. Since inflation is held near the saddle point, we can approximate the inflaton energy to be nearly constant, $V(\phi_0)$. The curvaton's motion can be solved by integrating out its slow roll equation of motion
\eq{\int_{e}^{*}\frac{\dd\sigma}{\sigma'}=\int_e^* \dd N,\label{eq-c-int1}}
where
\eq{\sigma'\equiv\frac{\partial\sigma}{\partial N}=-\frac{\partial\sigma}{H\partial t}=-\frac{U'(\sigma)}{3H^2}\,.}

After inflation, the inflaton decays into radiation,  whose energy density then satisfies
\eq{\rho_\gamma \approx V(\phi_0)e^{-4N_2},\label{eq-c-rg}}
where $N_2$ is the number of e-folds of universe expansion after inflation, till the curvaton $\sigma$ decays.

From the violation of the second order slow roll condition, i.e.\ $\eta_{\sigma c}=-1$, we find the curvaton slowly rolls after inflation for the number of e-folds $N_2$, which satisfies
\eq{\sigma_c=\sigma_0-\Delta \sigma e^{-4N_2},\label{eq-c-sc1}}
where
\eq{\Delta \sigma \equiv\frac{3H^2}{U'''(\sigma_0)}}
characterizes the typical ``width'' of the slow roll region of the curvaton field near the saddle point.

The slow roll equation of motion for the curvaton can also be integrated out after inflation, although now the universe is dominated by radiation. Similar to \refeq{c-int1}, here we have
\eq{\int_{e}^{c}\frac{\dd\sigma}{\sigma'}=\int_e^c\dd N\label{eq-c-int2}}
Therefore \refeq{c-int1} and \refeq{c-int2} fully describe the motion of the curvaton, before it ends slow roll. Since, $m_\sigma\gg H_*$ after the end of slow roll, $\sigma$ decays into radiation and the universe then evolves adiabatically~\footnote{The saddle point keeps the effective mass near the saddle point vanishing, but the bare mass term of the curvaton is of the order of the soft SUSY breaking term, i.e. $\sim 1-10$~TeV.}.

When the curvaton receives the quantum fluctuations $\delta\sigma_*$ at the Hubble exits, its initial perturbation will be maintained during inflation. Its perturbation evolves according to the perturbed \refeq{c-int1}, which is
\eq{\frac{\delta\sigma_*}{\sigma_*'}-\frac{\delta\sigma_e}{\sigma_e'}=0.\label{eq-c-d1}}

After inflation, the curvaton's perturbations are converted into the curvature perturbations when the curvaton decays \footnote{In fact in our case it happens soon after the curvaton ends its slow roll condition, by virtue of $m_\sigma \gg H_*$.}. This can be specified by the perturbation in the number of e-folds of the curvaton's slow roll after inflation, $\delta N_2$, and the perturbation in the curvaton field at the end of slow roll, $\delta\sigma_c$. Together they should comply the same end-of-slow-roll condition for the curvaton, i.e.\ $\delta\eta_{\sigma c}=0$, which gives
\eq{\frac{\delta\sigma_c}{\sigma_c-\sigma_0}=-4\delta N_2.\label{eq-c-d2}}

\fig{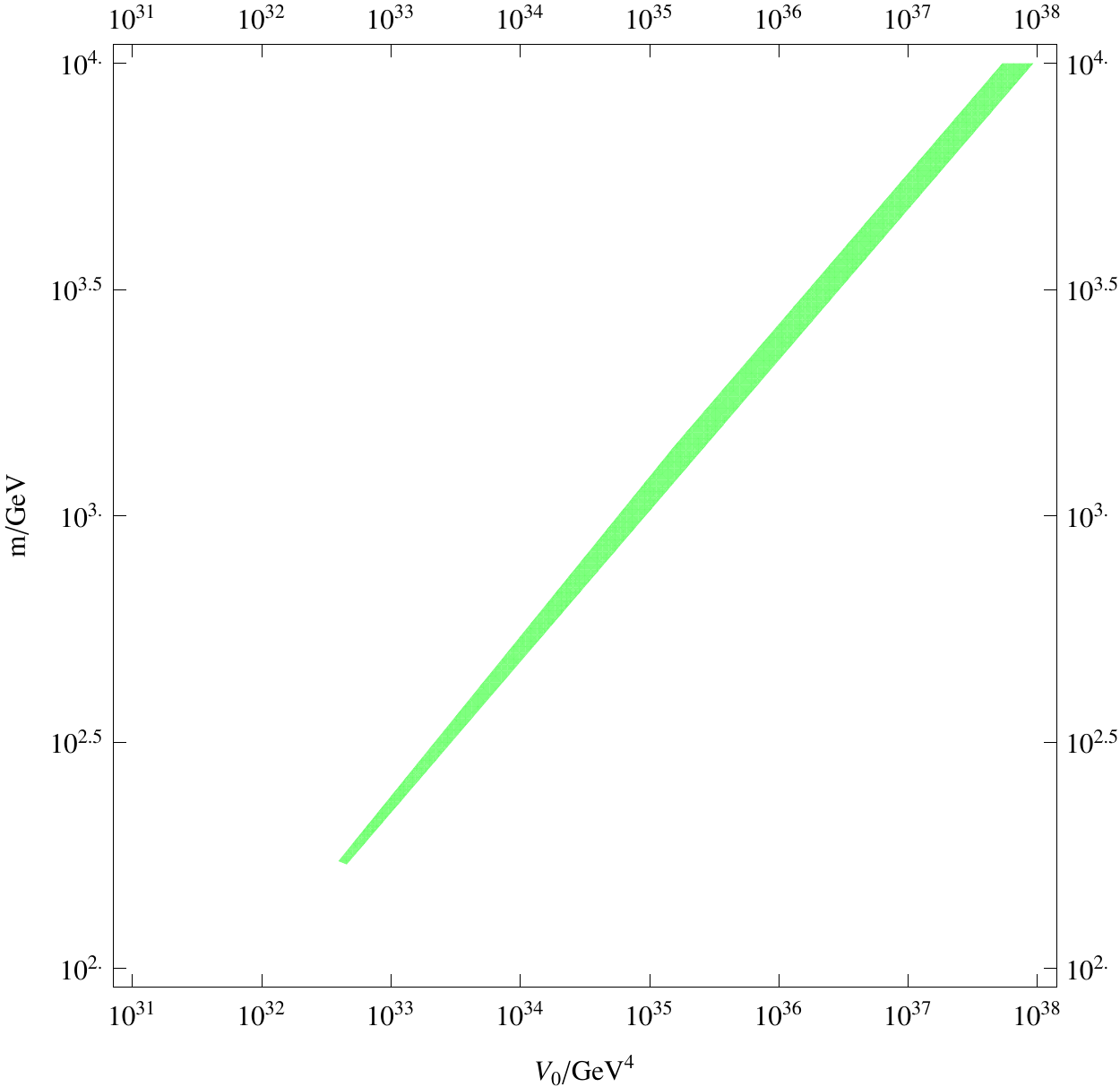}{Parameter space is scanned for the curvaton model for $\ms=5\times10^{16}$~GeV. The allowed band is painted in green, giving a positive local bispectrum within $\fNL=2.7\pm17.4$, the latest Planck observational constraint for $3\sigma$. The initial condition for the curvaton $\sigma_*$ and the coupling constant $\lambda$ have been picked (according to \refeq{c-Pz} and \refeq{c-sx}) to always match the observed central values for the power spectrum $P_\zeta=2.196\times10^{-9}$ and spectral index $n_s=0.9603$~\cite{Planckc}.}{0.8}{}

The curvaton perturbation, $\delta\sigma$, also evolves after the end of inflation, according to the perturbed \refeq{c-int2}, with the relation
\eq{\frac{\delta\sigma_c}{\sigma_c'}-\frac{\delta\sigma_e}{\sigma_e'}=\delta N_2.\label{eq-c-d3}}
Combining \refeq{c-d1}, \refeq{c-d2} and \refeq{c-d3}, we are able to solve the perturbation in the number of e-folds before the curvaton ends slow roll
\eq{\delta N_2=\frac{\delta\sigma_*}{7\sigma_*'}\label{eq-c-dN2}}

After the curvaton ends slow roll, the $\sigma$ field instantly decays into relativistic {\it dof} within one Hubble time and we can ignore its evolution. The e-folds of the radiation dominated era, $N_3$, can be written as a constant plus a quarter of the logarithmic of the total energy density, i.e.
\eq{N_3=\frac{1}{4}\log(\rho_\gamma+U_c)+const.\label{eq-c-N3}}
When the initial perturbation $\delta\sigma_*$ is present, it also changes the energy density at the time the curvaton ends slow roll. The perturbation in $\sigma$'s energy density, in this case, would be small compared to that of the radiation. This is both because the curvaton is subdominant, and because its potential is relatively flat around the saddle point. Therefore, the major contribution to $\delta N_3$ comes from the perturbation in the radiation energy density, which comes from perturbing \refeq{c-N3} as (according to \refeq{c-rg})
\eq{\delta N_3=-\frac{\rho_\gamma}{\rho_\gamma+U_c}\delta N_2=-(1-r)\delta N_2\,,}
where
\eq{r\equiv\frac{U_c}{\rho_\gamma+U_c}\approx\frac{U(\sigma_0)}{V(\phi_0)}e^{4N_2}\label{eq-c-r}}
is the energy density ratio of the curvaton when it decays.

After taking into account the energy density perturbation at the curvaton's end of slow roll, the universe then enters an adiabatic evolution, and no super Hubble perturbations will be generated. Therefore the total perturbation in the e-folds of expansion is
\eq{\delta N=\delta N_2+\delta N_3=N_\sigma\delta\sigma_*,}
where
\eq{N_\sigma=\frac{r}{7\sigma_*'}.}

Therefore the power spectrum of curvature perturbation becomes
\eq{P_\zeta=N_\sigma^2P_{\delta\sigma_*}=\left(\frac{r\Delta \sigma H_*}{7\pi(\sigma_*-\sigma_0)^2}\right)^2.\label{eq-c-Pz}}
The spectral index of the curvature perturbation can then be calculated 
 as
\eq{n_s-1\equiv\frac{\dd\ln P_\zeta}{\dd\ln k}=-\frac{2(\sigma_0-\sigma_*)}{\Delta \sigma }.}
From the above equations, we find that the observed spectral index $n_s$ constrains the initial condition of the curvaton $\sigma_*$ by
\eq{\sigma_*=\sigma_0+\frac{n_s-1}{2}\Delta \sigma.\label{eq-c-sx}}
The local bispectrum can be calculated, according to \refeq{c-dN2} and \refeq{c-r}, as
\eq{\fNL=\frac{5N_{\sigma\sigma}}{6N_\sigma^2}\approx\frac{5}{6N_\sigma}\frac{\partial\ln r}{\partial\sigma_*}=\frac{5\delta N_2}{6N_\sigma\delta\sigma_*}=\frac{5}{6r}\sim\frac{1}{r}.\label{eq-c-fNL}}

We can scan the parameter space in $m_\sigma$ and $V(\phi_0)$, and calculate the possible local bispectrum, as shown in \refig{C-PS}.

In \refig{C-PS}, we require the coupling $\lambda<1$, and the curvaton remains subdominant ($r<0.1$). These two constraints narrow the allowed parameter space to the green band. The local bispectrum typically acquires $\fNL\sim {\cal O}(20)$. As a specific example, we pick the inflaton as having the parameters $\lambda_2=10^{-9}$ and $\lambda_4=10^{-3}$, with an inflation energy scale relatively low, at $V(\phi_0)=(2.4\times10^9\,\mathrm{GeV})^4$, with negligible curvature perturbations. A low scale inflation is helpful in obtaining the right tilt in the power spectrum for the curvature perturbations. For the curvaton mass $m_\sigma=7.4\,\mathrm{TeV}$, we can acquire the observed power spectrum of curvaton perturbation by taking the coupling constant $\lambda=0.012$. The energy density ratio when curvaton decays is $r=0.086$. This gives the local bispectrum $\fNL=8.3$.


\section{Spectator mechanism with a visible sector inflation\label{sec-spectator}}

The spectator mechanism is a new mechanism, which has been proposed recently in Refs.~\cite{Mazumdar:2012rs,Wang:2013oea}. The perturbations are created by a sub-dominant field which decays during inflation, known as a spectator. The spectator field cannot modify the dominant inflaton dynamics, but it can leave its imprint in the cosmological perturbations. The decay of the spectator field can create non-Gaussianity due to the conversion of the entropy perturbations into the curvature perturbations. The spectator field decays completely into relativistic species during inflation, thus leaving no residual isocurvature perturbations. All the matter is created by the decay of the inflaton field after inflation, therefore it is important that the inflaton sector must be embedded within a well motivated visible sector. For all practical purposes the inflaton field's perturbations could be assumed to be sub dominant as compared to that of the spectator's. 

In order to create the right thermal history, we provide a simple example of embedding inflation within MSSM$\times U(1)_{B-L}$ gauge group, where the latter is also gauged. A simple $D$-gauge invariant flat direction which can be the inflaton candidate in our case is given by~\cite{Allahverdi:2006cx,B-L}:
\begin{equation}
W\supset h{\bf NH_uL}\,,
\end{equation}
where $h$ is the Yukawa coupling and, ${\bf N}$, ${\bf H_u}$, ${\bf L}$ are corresponding right handed neutrino, Higgs and slepton superfields. Note that the above superpotential can generate Dirac mass for the light neutrinos if the scale of $U(1)_{B-L}$ breaking is of order ${\cal O}(\rm TeV)$ and the Yukawa is $h\sim 10^{-11} -10^{-12}$.

The inflaton field $\phi$ corresponds to the superpotential:
\begin{equation}
\phi=\frac{{\widetilde N}+{H_u}+{\widetilde L}}{\sqrt{3}}\,.
\end{equation}
So its potential can be written as
\begin{equation}
V(\phi)=\frac{1}{2}m_{\phi}^2|\phi|^2-\frac{Ah}{6\sqrt{3}}\phi^3+\frac{h^2}{12}|\phi|^4\,,
\end{equation}
where $A$ is the trilinear A-term and the soft SUSY breaking mass term for the flat direction is given by:
\begin{equation}
m_{\phi}^2=\frac{m^2_{\widetilde N}+m^2_{H_u}+m^2_{\widetilde L}}{3}\,,
\end{equation}
Note that for $A = 4 m_{\phi}$, there exists a {\it saddle point} for
which $V'(\phi_0) = V''(\phi_0) = 0$. The saddle
point and the potential are given by:
\begin{eqnarray} \label{eq-s-sad} \phi_0 = \sqrt{3}\frac{m_{\phi}}{h}=
6 \times 10^{12} ~ m_{\phi} ~ \Big({0.05 ~
{\rm eV} \over m_\nu} \Big)\,, \\
\label{eq-s-sadpot}
V(\phi_0) = \frac{m_{\phi}^4}{4h^2}=3 \times 10^{24} ~ m^4_{\phi} ~
\Big({0.05 ~ {\rm eV} \over m_\nu} \Big)^2 \,.
\end{eqnarray}
Here $m_\nu$ denotes the neutrino mass which is given by $m_\nu = h
\langle H_u \rangle$, with $\langle H_u \rangle \simeq 174$ GeV. For
neutrino masses with a hierarchical pattern, the largest neutrino mass is $m_\nu \simeq 0.05$ eV in order to explain the atmospheric neutrino oscillations~\cite{atmos}. In our case we will be investigating a range of the inflaton masses which can accommodate the right handed sneutrino mass close to the low scale supersymmetry, and also yield the correct neutrino masses:
\begin{equation}
m_{\phi}\sim 1~{\rm TeV}~~(h\leq 10^{-11})\hspace{0.2in} \mathrm{to}\hspace{0.2in} m_{\phi}\sim 100~{\rm TeV}~~(h\leq 10^{-12})\,.
\end{equation}
The above range of $h$ also guarantees the curvature perturbation contribution by the inflaton is negligible, see \cite{Wang:2013oea}.

After inflation the inflaton $NH_uL$ will start coherent oscillations and will dump all its energy into the light relativistic species of MSSM and the lightest of the right handed sneutrinos. In fact the lightest right handed sneutrino could be the dark matter candidate if it is the lightest SUSY particle. Dark matter analysis with the lightest sneutrino has been performed in Ref.~\cite{B-L}.  Since $U(1)_{B-L}$ is gauged, it will lead to a quick thermalization of all the relativistic {\it dof} with a reheat temperature very similar to the analysis of Ref.~\cite{Allahverdi:2011aj}. The reheat temperature will be roughly given by $T_{rh}\sim (30/\pi^2g_\ast)^{1/4}V(\phi_0)^{1/4}\sim 10^{8}$~GeV for $m_{\phi}\sim 1$~TeV~\cite{B-L,Allahverdi:2011aj}. 

Let us now imagine that the spectator field has a simple potential arising from some hidden sector physics. For the sake of illustration we consider this to have a flat potential with a hyperbolic tangent profile:
\begin{equation}
U(\sigma) = \frac{U_0}{2}\left(1+\tanh\frac{\sigma}{\sigma_0}\right)\,.
\end{equation}
where $U_0$ and $\sigma_0$ are constant parameters whose values we will scan to show how this simple potential can explain the amplitude of the perturbations and also the observable non-Gaussianity of the {\it local } form.  Note that 
$U_0 \ll V(\phi_0)$.

A typical timeline for the spectator scenario in our case is summarized in \refig{S-Timeline}. We will use the subscripts ``$*$'', ``$c$'' and ``$e$'' to indicate the respective slices as the relevant perturbations from the spectator field leave the Hubble patch during inflation, the spectator ends slow roll, and the end of inflation.


\fig{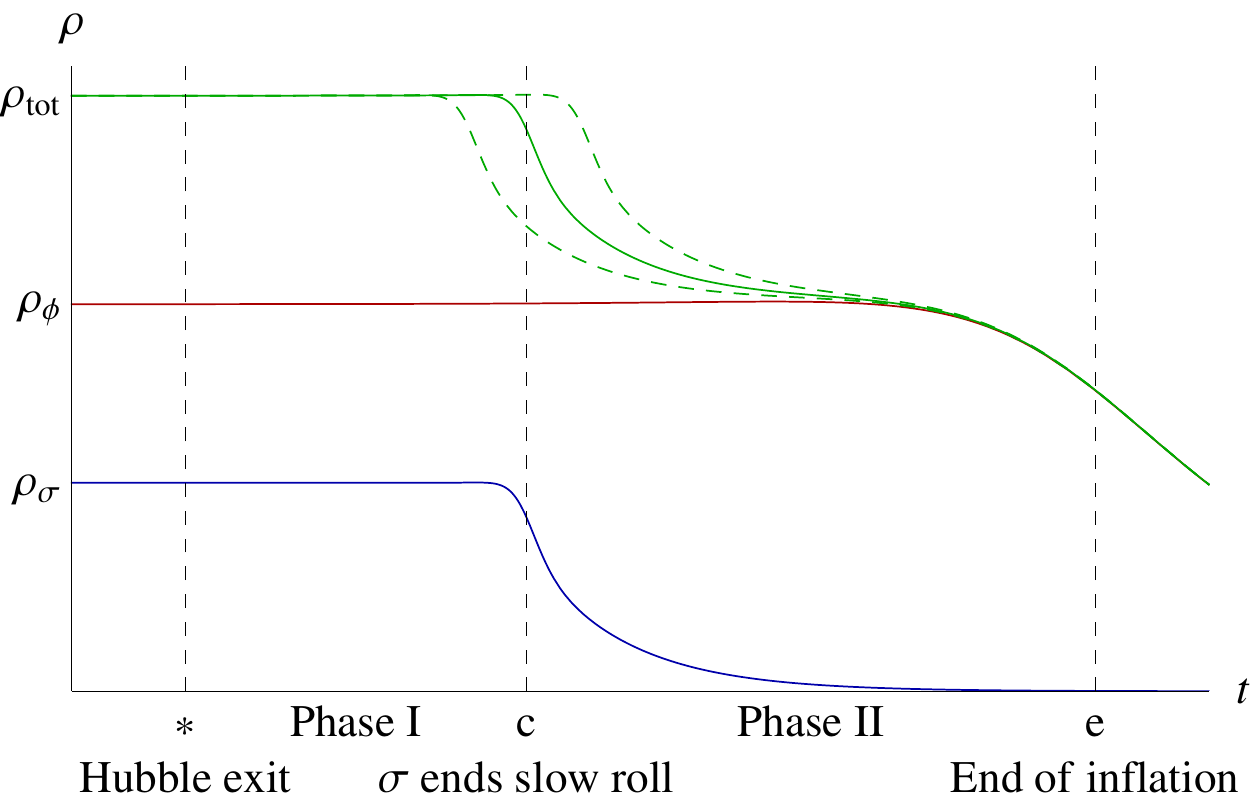}{A schematic timeline for the spectator scenario is shown. Two phases during inflation have been shown after the relevant modes have left the Hubble patch. The energy densities of the inflaton, the curvaton, and the total are drawn in red, blue and green respectively. The dashed green curves show how a perturbation in the spectator field may affect the universe's evolution.}{0.8}{}


\figa{\figi{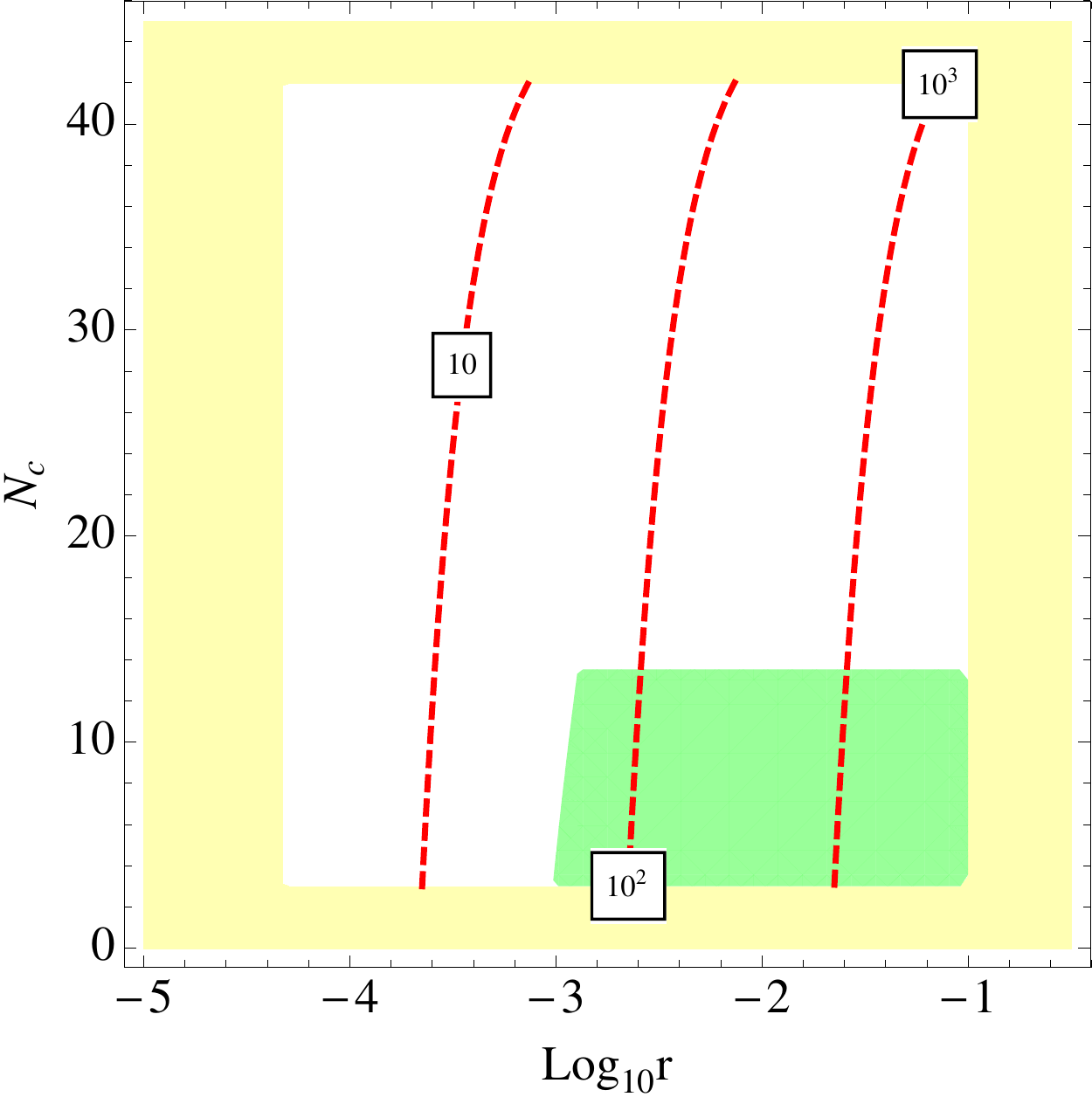}{The relative value $\sigma_0/H_*$.}{0.48}\hspace{0.04\textwidth}\figi{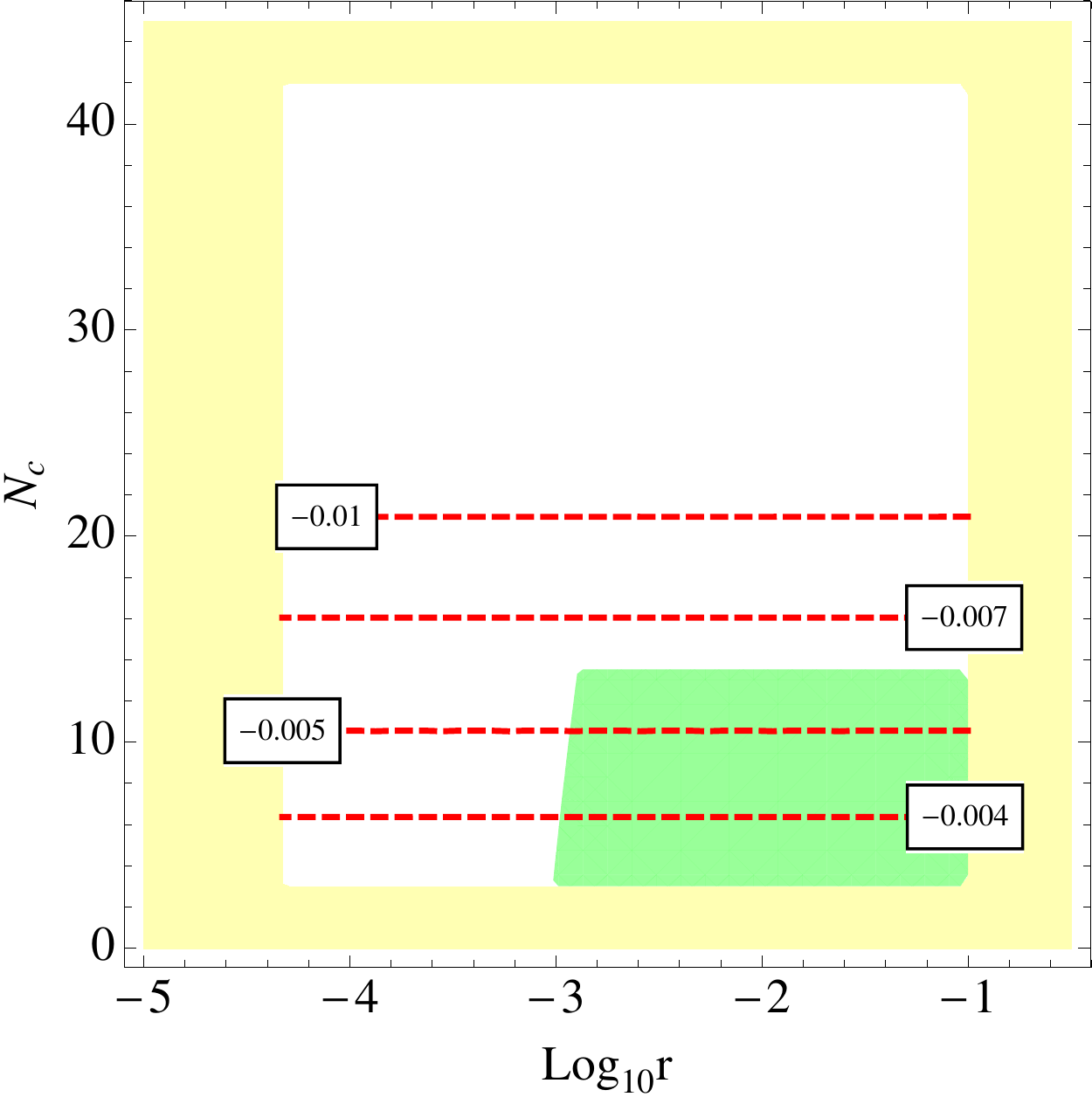}{The running of spectral index lies inside the current observational bound $\dd n_s/\dd\ln k=-0.0134\pm0.0270$ ($3\sigma$)~\cite{Planckc}.}{0.48}\\

\figi{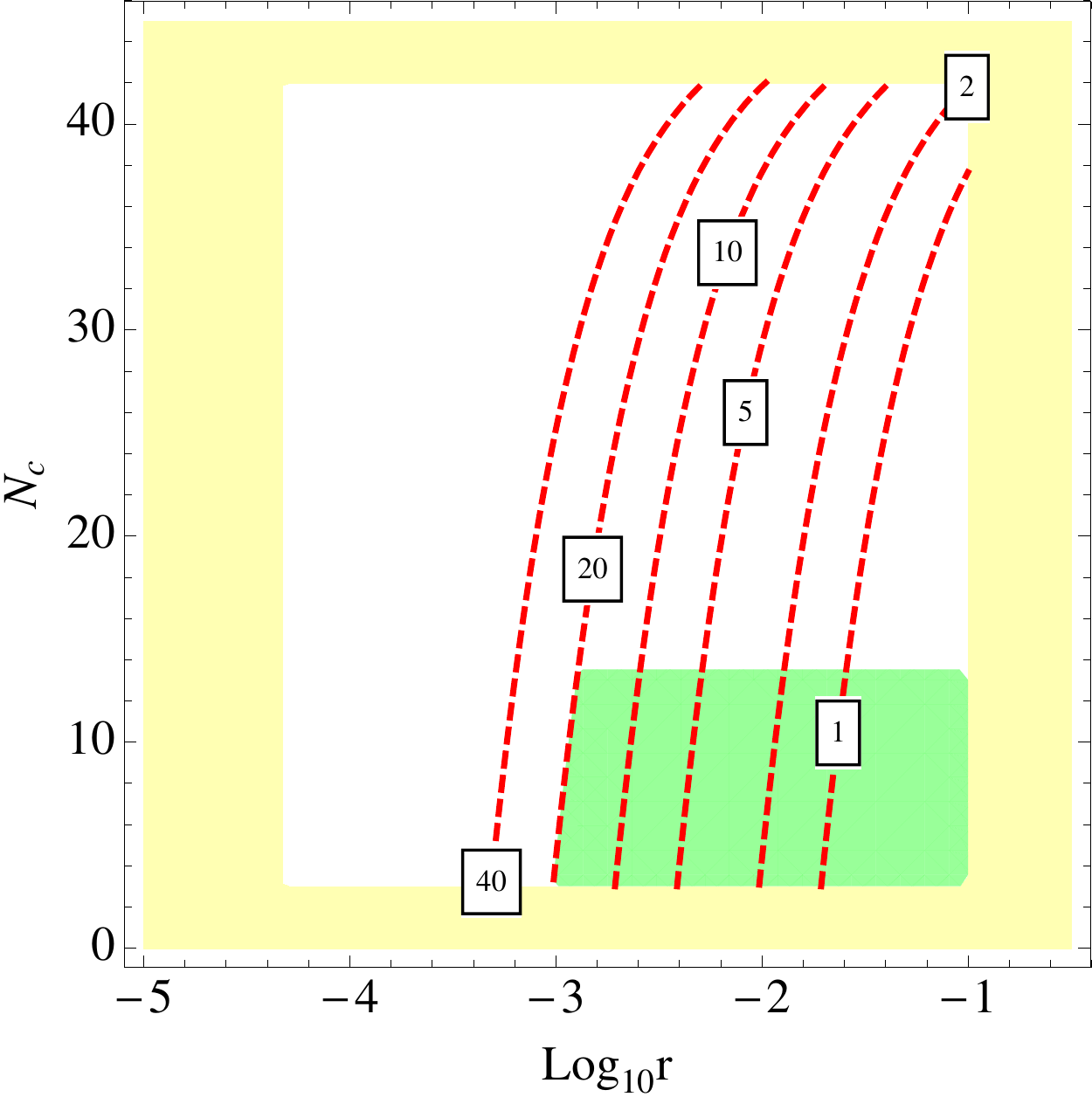}{The local bispectrum of curvature perturbations $\fNL$ gives a broad parameter space surviving from the Planck observational constraints~\cite{Planckng}.}{0.48}\hspace{0.04\textwidth}\figi{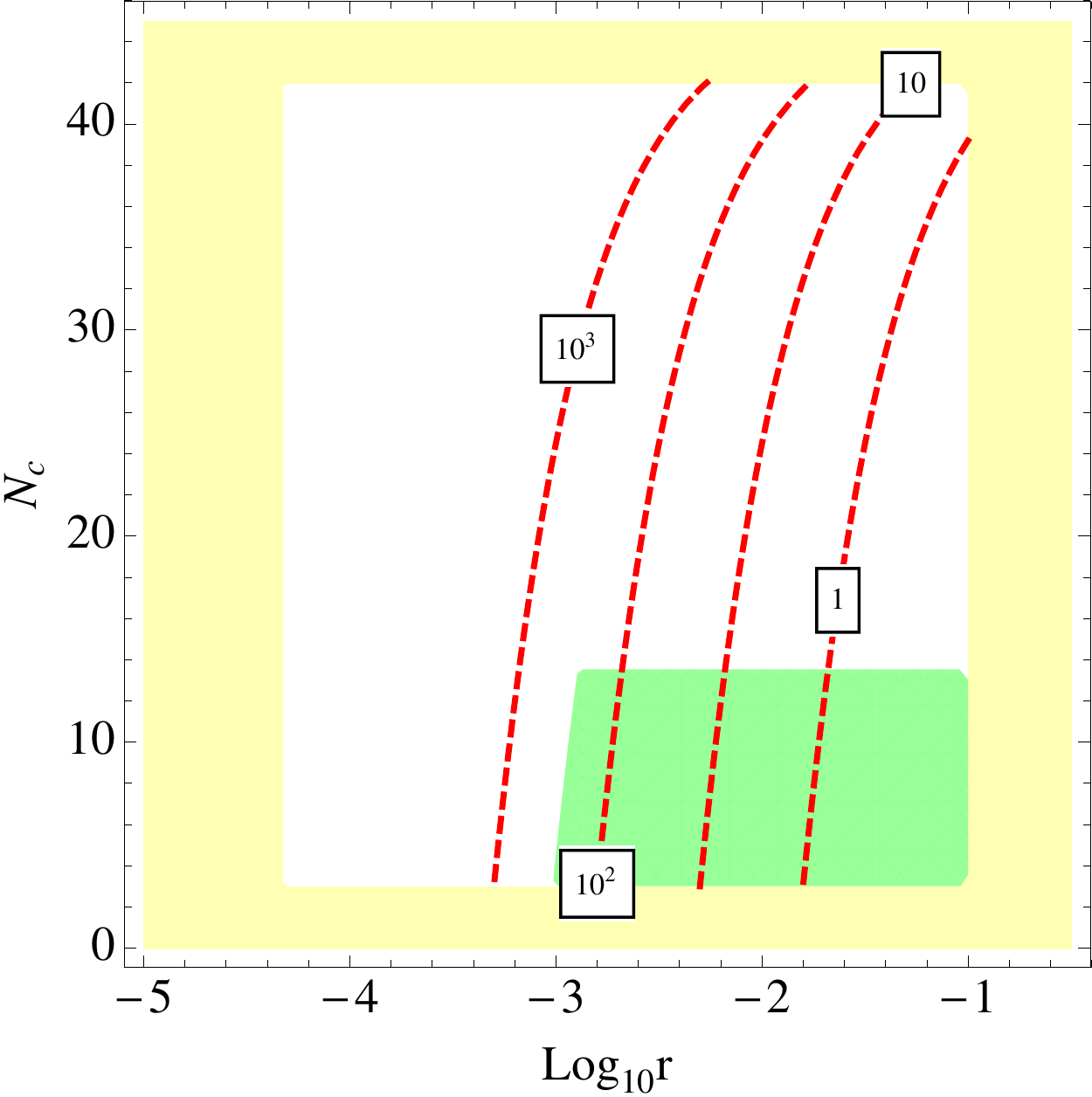}{The local trispectrum of curvature perturbations $\gNL$.}{0.48}}{The cosmological parameters for the spectator model. The yellow shaded regions are excluded due to multiple constraints. The green bands lie with the parameter space that gives a spectral index inside $n_s=0.9603\pm0.219$, and the local $\fNL=2.7\pm17.4$ from the Planck observation ($3\sigma$)~\cite{Plancki,Planckng}. The red contour lines are for the values of the respective parameters. Here we have taken the pivot scale e-folding $N_*=45$.\label{fig-S-params}}{}


We can solve the background evolution of the spectator field $\sigma$, typically $\sigma$ ends slow roll when
\eq{\eta_{\sigma c}\equiv\frac{\mpl^2U(\sigma_c)''}{V(\phi_0)}=-1.}
From this we can solve $\sigma_c$ as
\eq{e^{2\sigma_c/\sigma_0}=\frac{4\mpl^2U_0}{V(\phi_0)\sigma_0^2},}
for $\sigma_c>\sigma_0$. Here the subscript ``$c$'' indicates at the time when $\sigma$ ends slow roll. We can also solve the background motion of the spectator field $\sigma$ before it ends slow roll, since the inflaton is dominating with a constant energy density. By integrating the slow roll equation of motion for $\sigma$, we obtain $\sigma$, as a function of $N$, the remaining e-folds of inflation, as
\eq{e^{2\sigma/\sigma_0}=\frac{4\mpl^2U_0(N-N_c+1)}{V(\phi_0)\sigma_0^2},}
where $N_c$ is number of e-folds of inflation from $\sigma$ ends slow roll to the end of inflation. The second and third order slow roll parameters for $\sigma$ then simplify to
\begin{eqnarray}
\eta_{\sigma} &=&\frac{\mpl^2U''}{V(\phi_0)}=-\frac{1}{N-N_c+1}\,,\\
\xi_{\sigma} &=&\frac{\mpl^4U'U'''}{V(\phi_0)^2}=\frac{1}{(N-N_c+1)^2}\,.
\end{eqnarray}
For the pivot scale $N=N_*$, the spectral index $n_s$, the local bispectrum $\fNL$, and the local trispectrum $\gNL$ are determined by $\eta_{\sigma*}$ in this case, giving the leading order terms, see for 
details~\cite{Mazumdar:2012rs,Wang:2013oea}
\eqa{n_s-1&=&2\eta_{\sigma*}=-\frac{2}{N_*-N_c+1},\\
\fNL&=&-\frac{5\eta_{\sigma*}}{6r}=\frac{5}{6(N_*-N_c+1)r},\label{eq-s-fNL}\\
\gNL&=&\frac{25(2\eta_{\sigma*}^2-\xi_{\sigma*})}{54r^2}=\frac{25}{54(N_*-N_c+1)^2r^2}.}
where
\begin{equation}\label{eq-s-spectator-r}
r\equiv \frac{U(\sigma_c)}{(V(\phi_c)+U(\sigma_c))}\approx \frac{U_0}{V(\phi_0)}\,,
\end{equation}
is the energy density ratio of the spectator $\sigma$ at its end-of-slow-roll boundary.

The power spectrum of curvature perturbations then, becomes
\eq{P_\zeta=\frac{H_*^2}{4\pi^2}\left(\frac{U_*}{\mpl^2U_*'}\right)^2=\frac{(N_*-N_c+1)^2r^2V_0}{3\pi^2\mpl^2\sigma_0^2}.}
Therefore to achieve the observed amplitude for $P_\zeta$, this requires $\sigma_0$ to take the value
\eq{\frac{\sigma_0^2}{H_*^2}=\frac{(N_*-N_c+1)^2r^2}{\pi^2P_\zeta}.}

When the inflaton model is given, the parameters $N_*$ and $V_0$ are fixed. The spectral index $n_s$, the local bispectrum $\fNL$, the local trispectrum $\gNL$, and the relative value $\sigma_0/H_*$ then only depend on the $r$, the energy density ratio, and $N_c$, the number of e-folds from the spectator ends slow roll to the end of inflation. We can then parametrically plot their dependences on $N_c$ and $r$ in \refig{S-params}.

One can see that the model predicts the spectral tilt, (as shown in green  shade which depicts the $2\sigma$ range,) the negligible running of the spectral tilt and the local bispectrum in the range observed by the current Planck data~\cite{Planckc,Planckng,Plancki}. We have also shown the value of $g_{NL}$ in \refig{S-params}.  Since $\sigma$ decays into radiation there is no residual isocurvature fluctuations, which matches the data perfectly well.

Since the origin of $\sigma$ field belongs to the hidden sector, it could arise from billions of hidden sectors of string theory, all the inflationary models which claim a successful inflationary cosmology could potentially act like a spectator field. One of the common feature for the spectator field is the flat potential and this can be achieved  in many string motivated models. Now the advantage is that these stringy origins need not have to explain the matter content of the universe. The latter could be obtained from the visible sector model of inflation.


\secs{Nonlocal bispectra from curvaton and spectator mechanisms}

So far we have discussed the non-Gaussianity of local type or in the squeezed limit. In this section we briefly discuss how to obtain the equilateral and orthogonal types of the bispectrum. The bispectrum for any shape $\fNL(\vect k_1,\vect k_2)$ is in general defined as~\cite{Bartolo}
\eq{B(\vect k_1,\vect k_2,\vect k_3)=\frac{6}{5}\fNL(\vect k_1,\vect k_2)(P(\vect k_1)P(\vect k_2)+P(\vect k_2)P(\vect k_3)+P(\vect k_3)P(\vect k_1)).\label{eq-nl-fNL0}}
Here $B$ and $P$ indicate the strengths of the two-and three-point correlation functions of the gauge invariant curvature perturbation $\zeta$. They are defined as
\eq{\langle\zeta(\vect k_1)\zeta(\vect k_2)\rangle=8\pi^3P(\vect k_1)\delta^3(\vect k_1+\vect k_2),}
\eq{\langle\zeta(\vect k_1)\zeta(\vect k_2)\zeta(\vect k_3)\rangle=8\pi^3B(\vect k_1,\vect k_2,\vect k_3)\delta^3(\vect k_1+\vect k_2+\vect k_3),\label{eq-nl-B}}

In \refeq{nl-B}, the delta function indicates $\vect k_1$, $\vect k_2$, and $\vect k_3$ form a closed momentum triangle, because of this multi-point correlation function expansion. Depending on the shape of the triangle, the bispectrum $\fNL(\vect k_1,\vect k_2)$ can take different values, corresponding to different types of the bispectrum. The most considered and also best observationally constrained type of bispectrum, the ``local'' type which we have discussed above, corresponds to the squeezed limit $k_1\approx k_2\gg k_3$. For any single field slow roll inflation, the local bispectrum of curvature perturbations is constrained to be small by the almost scale invariant spectrum~\cite{Maldacena:2002vr}.

In principle, every other shape of the momentum triangle corresponds to a unique type of bispectrum, and is considered ``nonlocal''. Among them, however, what raises people's most interests are the ``equilateral'' and ``orthogonal'' types. The equilateral type comes from taking the equilateral momentum triangle where $k_1=k_2=k_3$, and the excess equilateral bispectrum (compared to the local one) typically can be generated by models with non-canonical kinetic terms~\cite{noncanonical,Senatore:2009gt}. The orthogonal type is constructed in \cite{Senatore:2009gt}, to account for the bispectrum contribution that is orthogonal/uncorrelated to the local and equilateral shapes, for the general single field inflation. For this reason, the orthogonal bispectrum is in general a linear combination of various shapes, and does not originate from a single-shape definition.

It has been shown in the past that the non-canonical inflatons can generate nonlocal bispectra with various patterns~\cite{noncanonical}. If we embed such a non-canonical field in the de Sitter universe, this non-canonical field can still contribute to the nonlocal bispectra as a curvaton or a spectator~\cite{Cai:2010rt}. The inflaton field dominating the de Sitter universe can still come from the visible sector, as demonstrated in \refsec{curvaton} and \refsec{spectator}, to be responsible for the matter production.

The generic nonlocal bispectrum for the curvature perturbation $\zeta$ for any nonlocal shape, written as $\fNL\sups{(nloc)}(\vect k_1,\vect k_2)$, can be estimated as follows. In both the curvaton and the spectator cases, the universe evolves almost adiabatically till the boundary where the curvaton or spectator changes its equation of state significantly. Before this boundary, the gauge invariant perturbation of the curvaton or spectator, which is defined as
\eq{\zeta_\sigma(\vect k)=-\psi(\vect k)-H\frac{\delta\rho_\sigma(\vect k)}{\rho_\sigma'},}
does not evolve after the Hubble exit of the relevant modes. Here $\sigma$ is the curvaton or the spectator field we are concerned about, $\rho_\sigma$ is its energy density, $\psi(\vect k)$ is the scalar perturbation in the metric, and here the prime means derivative w.r.t the conformal time. We have omitted the time dependence.

Here we can define the nonlocal bispectrum for the gauge invariant perturbation $\zeta_\sigma(\vect k)$, as $f_{\mathrm{NL}(\sigma)}\sups{(nloc)}(\vect k_1,\vect k_2)$, similarly with \refeq{nl-fNL0} to \refeq{nl-B}~\footnote{It should be noted that here $f_{\mathrm{NL}(\sigma)}\sups{(nloc)}(\vect k_1,\vect k_2)$ is the nonlocal bispectrum of $\sigma$ in the de Sitter universe dominated by an inflaton field, as opposed to $\sigma$'s nonlocal bispectrum when itself serves as the inflaton. Therefore, typically the $\zeta_\sigma$ or $f_{\mathrm{NL}(\sigma)}\sups{(nloc)}(\vect k_1,\vect k_2)$ of the general single field inflation, may be applied as those in this discussion, but only after a scaling relation (as a function of $r$).}. In particular, we have
\eq{\langle\zeta_\sigma(\vect k_1)\zeta_\sigma(\vect k_2)\rangle=8\pi^3P_\sigma(\vect k_1)\delta^3(\vect k_1+\vect k_2),}
\eq{\langle\zeta_\sigma(\vect k_1)\zeta_\sigma(\vect k_2)\zeta_\sigma(\vect k_3)\rangle=8\pi^3B_\sigma(\vect k_1,\vect k_2,\vect k_3)\delta^3(\vect k_1+\vect k_2+\vect k_3),\label{eq-nl-Bs}}
\eq{B_\sigma(\vect k_1,\vect k_2,\vect k_3)=\frac{6}{5}f_{\mathrm{NL}(\sigma)}\sups{(nloc)}(\vect k_1,\vect k_2)(P_\sigma(\vect k_1)P_\sigma(\vect k_2)+P_\sigma(\vect k_2)P_\sigma(\vect k_3)+P_\sigma(\vect k_3)P_\sigma(\vect k_1)).\label{eq-nl-fNLs0}}

At the spectator or the curvaton boundary, the quantum fluctuations in $\sigma$ then transfer to the curvature perturbations. In the simplest setup where we can assume that the inflaton contribution to the curvature perturbation is negligible, the total curvature perturbation becomes
\eq{\zeta(\vect k)=r\zeta_\sigma(\vect k)\,,\label{eq-nl-zeta}}
where $r$ is the energy density ratio of $\sigma$ compared to the total energy density when it reaches the boundary, i.e. at the time when the perturbations are converted.

After the boundary, the spectator is quickly redshifted away and the curvaton decays into radiation which has the same equation of state with the inflaton's decay products. The universe then becomes adiabatic and $\zeta(\vect k)$ does not evolve afterwards. As the observation confirms, we can assume the local bispectrum, determined by \refeq{c-fNL} or \refeq{s-fNL}, is much smaller than the nonlocal bispectrum we are interested in. Therefore the nonlocal bispectrum of the curvature perturbations can then be calculated, according to \refeq{nl-zeta}, as
\eqa{\fNL\sups{(nloc)}(\vect k_1,\vect k_2)&=&\left.\frac{5B(\vect k_1,\vect k_2,\vect k_3)}{6(P(\vect k_1)P(\vect k_2)+\cdots)}\right|_{\vect k_3=-\vect k_1-\vect k_2}\nonumber\\
&=&\frac{1}{r}\left.\frac{5B_\sigma(\vect k_1,\vect k_2,\vect k_3)}{6(P_\sigma(\vect k_1)P_\sigma(\vect k_2)+\cdots)}\right|_{\vect k_3=-\vect k_1-\vect k_2}\nonumber\\
&=&\frac{1}{r}f_{\mathrm{NL}(\sigma)}\sups{(nloc)}(\vect k_1,\vect k_2)\,.\label{eq-nl-fNLnl}}
The above equation also works for both the equilateral and the orthogonal types of bispectra.

The multi-point correlation functions has been well studied for a noncanonical slow roll field in a de Sitter universe, showing possible large equilateral and/or orthogonal types of bispectrum~\cite{noncanonical}. The \refeq{nl-fNLnl} then gives a simple enhancing relation for the nonlocal bispectrum of curvature perturbations, showing that the same mechanism also works in the curvaton or the spectator setup, in order to generate large nonlocal bispectra. 

Typically non-canonical kinetic terms such those as appearing in the DBI-inflation occur in the hidden sector rather than the visible sector particle physics. It is simple to hide the uncertainties of the hidden sector physics in the spectator scenario as compared to the visible sector curvaton. One can imagine a spectator field with a non-canonical kinetic term slowly rolling its potential and decaying well before the end of visible sector inflation. The decay products of the spectator field would be diluted away anyway during the remaining inflation, rendering the universe {\it solely} with the visible sector inflaton which produces all the relevant matter for the BBN and without any trace of dark radiation.

\secs{Summary}

In this paper we have emphasized only the visible sector models of inflation, curvaton  and spectator scenarios in light of Planck data~\cite{Planckc,Planckng,Plancki}. There are plethora of hidden sector gauge singlet models of inflation and curvaton with ad-hoc couplings and mass parameters, but these models are unable to explain why such a gauge singlet inflaton or curvaton would decay {\it solely} into the visible sector {\it dof}. We summarize our findings in Table-1. Although we have restricted ourselves to the local form of $\fNL$ but one advantage of a spectator scenario is that it can generate nonlocal forms of $\fNL$ compatible with the Planck data.

\begin{table}
\begin{center}
\caption{\label{tab:summary}Benchmark points considered in this study of MSSM inflation, MSSM curvaton and a spectator scenario with $MSSM\times U(1)_{B-L}$ flat direction inflaton. ``$\checkmark$'' means the model prediction satisfies the latest constraint by the Planck data. None of these models produce observable isocurvature perturbations.}
\vspace{0.2in}
\begin{tabular*}{0.97\textwidth}{|c|c|c|c|}
\hline Planck Constraints (1$\sigma$) & MSSM inflation & MSSM Curvaton & Spectator \\ \hline \hline
\parbox{1.7in}{\vspace{0.02in}\centering Tensor-to-scalar ratio\newline$r<0.11$ (95\% CL)~\cite{Plancki}}& Negligible, $\checkmark$ & Negligible, $\checkmark$ & Negligible, $\checkmark$\vspace{0.004in}\\ \hline
$10^9P_{\zeta}=2.196^{+0.051}_{-0.060}$~\cite{Planckc}& $\checkmark$ & $\checkmark$& $\checkmark$ \\ \hline
$n_s= 0.9603\pm0.073$~\cite{Planckc}& $\checkmark$ & $\checkmark$& $\checkmark$ \\ \hline
$dn_s/d\ln k=-0.0134\pm0.0090$~\cite{Plancki}& $\lesssim-0.002$, \checkmark & $\checkmark$& $\checkmark$ \\ \hline
$\fNL^{local}=2.7\pm5.8$~\cite{Planckng}&$<1$, \checkmark & Constrained, \checkmark&$\checkmark$\\ \hline
$\fNL^{equil}=-42\pm75$~\cite{Planckng}& $<1$, \checkmark & Constrained, \checkmark& $\checkmark$\\ \hline
$\fNL^{orth}=-25\pm39$~\cite{Planckng}&$<1$, \checkmark & Constrained, \checkmark & $\checkmark$ \\ \hline
Relativistic {\it dof}~\cite{Planckc}& only SM &  only SM&  only SM\\ \hline 
\end{tabular*}
\end{center}
\end{table}

The spectator mechanism is amazing since it can fit all the observed parameters, ${\cal P}_{\zeta}, n_s, dn_s/d\ln k, \fNL^{local},
\fNL^{oath}, \fNL^{equil}$ of the CMB without generating any {\it isocurvature} perturbations, besides providing all the SM {\it dof}, which means no dark radiation by virtue of the inflaton being within the visible sector physics.

\acknowledgments{}
The research of AM is supported by the Lancaster-Manchester-Sheffield Consortium for Fundamental Physics under STFC grant ST/J000418/1. EP is supported by STFC ST/J501074.


\begin{thebibliography}{1}


\bibitem{Guth:1980zm} 
  A.~H.~Guth,
  Phys.\ Rev.\ D {\bf 23}, 347 (1981).
  A.~D.~Linde,
  Phys.\ Lett.\ B {\bf 108}, 389 (1982).
  A.~Albrecht and P.~J.~Steinhardt,
  Phys.\ Rev.\ Lett.\  {\bf 48}, 1220 (1982).
  
  \bibitem{Planckc}
  http://www.sciops.esa.int/SA/PLANCK/docs/Planck\_2013\_results\_16.pdf
   \bibitem{Planckng}
  http://www.sciops.esa.int/SA/PLANCK/docs/Planck\_2013\_results\_22.pdf
  \bibitem{Plancki}
  http://www.sciops.esa.int/SA/PLANCK/docs/Planck\_2013\_results\_24.pdf
  
  \bibitem{Mazumdar:2010sa} 
  A.~Mazumdar and J.~Rocher,
  Phys.\ Rept.\  {\bf 497}, 85 (2011)
  [arXiv:1001.0993 [hep-ph]].

\bibitem{Allahverdi:2010xz} 
  R.~Allahverdi, R.~Brandenberger, F.~-Y.~Cyr-Racine and A.~Mazumdar,
  Ann.\ Rev.\ Nucl.\ Part.\ Sci.\  {\bf 60}, 27 (2010)
  [arXiv:1001.2600 [hep-th]].
  
  
  \bibitem{Burgess:2005sb} 
  C.~P.~Burgess, R.~Easther, A.~Mazumdar, D.~F.~Mota and T.~Multamaki,
  JHEP {\bf 0505}, 067 (2005)
  [hep-th/0501125].
  
  \bibitem{eternal}
  A.A. Starobinsky, in Current Topics in Field Theory, Quantum Gravity and Strings, eds. H.J. de Vega and N. Sanchez, Lecture Notes in Physics 206, Springer, Heidel- berg (1986); M. Miji?c, Phys. Rev. D 42, 2469 (1990)
  A.~D.~Linde, D.~A.~Linde and A.~Mezhlumian,
  Phys.\ Rev.\ D {\bf 49}, 1783 (1994)
  [gr-qc/9306035].
  
  \bibitem{linde-book}
   A.~D.~Linde,
  Contemp.\ Concepts Phys.\  {\bf 5}, 1 (1990)
  [hep-th/0503203].

\bibitem{Liddle:1998jc} 
  A.~R.~Liddle, A.~Mazumdar and F.~E.~Schunck,
  Phys.\ Rev.\ D {\bf 58}, 061301 (1998)
  [astro-ph/9804177].


\bibitem{Jokinen:2004bp} 
  A.~Jokinen and A.~Mazumdar,
  Phys.\ Lett.\ B {\bf 597}, 222 (2004)
  [hep-th/0406074].
  

\bibitem{strings-A}
A.~Mazumdar, S.~Panda and A.~Perez-Lorenzana,
  Nucl.\ Phys.\ B {\bf 614}, 101 (2001)
  [hep-ph/0107058].
  K.~Becker, M.~Becker and A.~Krause,
  Nucl.\ Phys.\ B {\bf 715}, 349 (2005)
  [hep-th/0501130].
   S.~Dimopoulos, S.~Kachru, J.~McGreevy and J.~G.~Wacker,
  JCAP {\bf 0808}, 003 (2008)
  [hep-th/0507205].
  

\bibitem{Allahverdi:2007zz} 
  R.~Allahverdi and A.~Mazumdar,
  Phys.\ Rev.\ D {\bf 76}, 103526 (2007)
  [hep-ph/0603244].
  

  \bibitem{Linde:2004zz} 
  A.~Linde,
  eConf C {\bf 040802}, L024 (2004).
  
  \bibitem{Polchinski:1998rr} 
  J.~Polchinski,
  Cambridge, UK: Univ. Pr. (1998) 531 p
  
  
  \bibitem{Cicoli:2010ha} 
  M.~Cicoli and A.~Mazumdar,
  JCAP {\bf 1009}, 025 (2010)
  [arXiv:1005.5076 [hep-th]].
   M.~Cicoli and A.~Mazumdar,
  Phys.\ Rev.\ D {\bf 83}, 063527 (2011)
  [arXiv:1010.0941 [hep-th]].
  
  \bibitem{Cicoli:2012aq} 
  M.~Cicoli, J.~P.~Conlon and F.~Quevedo,
  arXiv:1208.3562 [hep-ph].
  
  \bibitem{Chialva:2012rq} 
  D.~Chialva, P.~S.~B.~Dev and A.~Mazumdar,
  arXiv:1211.0250 [hep-ph].
  
  
 \bibitem{Pospelov:2010hj} 
  M.~Pospelov and J.~Pradler,
  Ann.\ Rev.\ Nucl.\ Part.\ Sci.\  {\bf 60}, 539 (2010)
  [arXiv:1011.1054 [hep-ph]].
 
   
  

  \bibitem{Chung:2003fi} 
  D.~J.~H.~Chung, L.~L.~Everett, G.~L.~Kane, S.~F.~King, J.~D.~Lykken and L.~-T.~Wang,
  Phys.\ Rept.\  {\bf 407}, 1 (2005)
  [hep-ph/0312378].
  
  \bibitem{ATLAS:2012ae} 
  G.~Aad {\it et al.}  [ATLAS Collaboration],
  Phys.\ Lett.\ B {\bf 710}, 49 (2012)
  [arXiv:1202.1408 [hep-ex]].


\bibitem{cms} 
  S.~Chatrchyan {\it et al.}  [CMS Collaboration],
  arXiv:1202.1488 [hep-ex].
 
  \bibitem{Feng:2013pwa} 
  J.~L.~Feng,
  arXiv:1302.6587 [hep-ph].
  
  \bibitem{Enqvist:2003gh} 
  K.~Enqvist and A.~Mazumdar,
  Phys.\ Rept.\  {\bf 380}, 99 (2003)
  [hep-ph/0209244].
  M.~Dine and A.~Kusenko,
  Rev.\ Mod.\ Phys.\  {\bf 76}, 1 (2003)
  [hep-ph/0303065].


  \bibitem{CDM}
G.~Jungman, M.~Kamionkowski and K.~Griest,
  Phys.\ Rept.\  {\bf 267}, 195 (1996)
  [hep-ph/9506380].
  
  
  \bibitem{david}
D.~H.~Lyth and D.~Wands,
  Phys.\ Lett.\  B {\bf 524} (2002) 5;
D.~H.~Lyth, C.~Ungarelli and D.~Wands,
  Phys.\ Rev.\  D {\bf 67} (2003) 023503.

\bibitem{enqvist}
K.~Enqvist and M.~S.~Sloth,
  Nucl.\ Phys.\  B {\bf 626} 395 (2002).

\bibitem{moroi}
T.~Moroi and T.~Takahashi,
  Phys.\ Lett.\  B {\bf 522} 215 (2001)
  [Erratum-ibid.\  B {\bf 539} 303 (2002)].
  
  
  \bibitem{Mazumdar:2012rs} 
  A.~Mazumdar and L.~Wang,
  arXiv:1210.7818 [astro-ph.CO].
  
\bibitem{Wang:2013oea} 
  L.~Wang and A.~Mazumdar,
  JCAP {\bf 1305}, 012 (2013)
  [arXiv:1302.2637 [astro-ph.CO]].

\bibitem{Bartolo}  
   N.~Bartolo, E.~Komatsu, S.~Matarrese and A.~Riotto,
  Phys.\ Rept.\  {\bf 402}, 103 (2004)
  [astro-ph/0406398].
  
\bibitem{Biswas:2013fna} 
  T.~Biswas, T.~Koivisto and A.~Mazumdar,
  arXiv:1302.6415 [astro-ph.CO].

\bibitem{Allahverdi:2006iq} 
  R.~Allahverdi, K.~Enqvist, J.~Garcia-Bellido and A.~Mazumdar,
  Phys.\ Rev.\ Lett.\  {\bf 97}, 191304 (2006)
  [hep-ph/0605035].


\bibitem{Allahverdi:2006we} 
  R.~Allahverdi, K.~Enqvist, J.~Garcia-Bellido, A.~Jokinen and A.~Mazumdar,
  JCAP {\bf 0706}, 019 (2007)
  [hep-ph/0610134].
  
  \bibitem{Allahverdi:2006cx} 
  R.~Allahverdi, A.~Kusenko and A.~Mazumdar,
  JCAP {\bf 0707}, 018 (2007)
  [hep-ph/0608138].


\bibitem{Mazumdar:2011xe} 
  A.~Mazumdar and S.~Nadathur,
  Phys.\ Rev.\ Lett.\  {\bf 108}, 111302 (2012)
  [arXiv:1107.4078 [hep-ph]].


\bibitem{Ashoorioon:2012kh} 
  A.~Ashoorioon, P.~S.~B.~Dev and A.~Mazumdar,
  arXiv:1211.4678 [hep-th].

\bibitem{Mukhanov:1990me} 
  V.~F.~Mukhanov, H.~A.~Feldman and R.~H.~Brandenberger,
  Phys.\ Rept.\  {\bf 215}, 203 (1992).


\bibitem{Gherghetta:1995dv} 
  T.~Gherghetta, C.~F.~Kolda and S.~P.~Martin,
  Nucl.\ Phys.\ B {\bf 468}, 37 (1996)
  [hep-ph/9510370].
  

\bibitem{Dine:1995kz} 
 M.~Dine, L.~Randall and S.~D.~Thomas,
  Nucl.\ Phys.\ B {\bf 458}, 291 (1996)
  [hep-ph/9507453].
  
\bibitem{Martin:2013tda} 
  J.~Martin, C.~Ringeval and V.~Vennin,
  arXiv:1303.3787 [astro-ph.CO].

  \bibitem{Allahverdi:2007vy} 
  R.~Allahverdi, B.~Dutta and A.~Mazumdar,
  Phys.\ Rev.\ D {\bf 75}, 075018 (2007)
  [hep-ph/0702112 [HEP-PH]].

\bibitem{Allahverdi:2007wh} 
  R.~Allahverdi, A.~R.~Frey and A.~Mazumdar,
  Phys.\ Rev.\ D {\bf 76}, 026001 (2007)
  [hep-th/0701233].

\bibitem{Allahverdi:2008bt} 
  R.~Allahverdi, B.~Dutta and A.~Mazumdar,
  Phys.\ Rev.\ D {\bf 78}, 063507 (2008)
  [arXiv:0806.4557 [hep-ph]].


\bibitem{Nilles:1983ge} 
  H.~P.~Nilles,
  Phys.\ Rept.\  {\bf 110}, 1 (1984).

\bibitem{ArkaniHamed:2004fb} 
  N.~Arkani-Hamed and S.~Dimopoulos,
  JHEP {\bf 0506}, 073 (2005)
  [hep-th/0405159].

\bibitem{Giudice:2004tc} 
  G.~F.~Giudice and A.~Romanino,
  Nucl.\ Phys.\ B {\bf 699}, 65 (2004)
  [Erratum-ibid.\ B {\bf 706}, 65 (2005)]
  [hep-ph/0406088].

\bibitem{Babu:2005ui} 
  K.~S.~Babu, T.~.Enkhbat and B.~Mukhopadhyaya,
  Nucl.\ Phys.\ B {\bf 720}, 47 (2005)
  [hep-ph/0501079].

\bibitem{Bueno Sanchez:2006xk} 
  J.~C.~Bueno Sanchez, K.~Dimopoulos and D.~H.~Lyth,
  JCAP {\bf 0701}, 015 (2007)
  [hep-ph/0608299].

\bibitem{Enqvist:2010vd} 
  K.~Enqvist, A.~Mazumdar and P.~Stephens,
  JCAP {\bf 1006}, 020 (2010)
  [arXiv:1004.3724 [hep-ph]].
  S.~Hotchkiss, A.~Mazumdar and S.~Nadathur,
  JCAP {\bf 1106}, 002 (2011)
  [arXiv:1101.6046 [astro-ph.CO]].


\bibitem{Mazumdar:2011ih} 
  A.~Mazumdar, S.~Nadathur and P.~Stephens,
  Phys.\ Rev.\ D {\bf 85}, 045001 (2012)
  [arXiv:1105.0430 [hep-th]].
  
  \bibitem{Hotchkiss:2011gz} 
  S.~Hotchkiss, A.~Mazumdar and S.~Nadathur,
  JCAP {\bf 1202}, 008 (2012)
  [arXiv:1110.5389 [astro-ph.CO]].
  
  \bibitem{Felder}
  G.~N.~Felder, L.~Kofman and A.~D.~Linde,
  Phys.\ Rev.\ D {\bf 59}, 123523 (1999)
  [hep-ph/9812289].

 \bibitem{Allahverdi:2011aj} 
  R.~Allahverdi, et. al,
  Phys.\ Rev.\ D {\bf 83}, 123507 (2011)
  [arXiv:1103.2123 [hep-ph]].
  
  \bibitem{Boehm:2012rh}
  C.~Boehm, J.~Da Silva, A.~Mazumdar and E.~Pukartas,
  Phys.\ Rev.\ D {\bf 87} (2013) 023529
  [arXiv:1205.2815 [hep-ph]].

  
\bibitem{mSUGRA}
A. H. Chamseddine, R. L. Arnowitt, P. Nath, Phys.Rev. Lett. {\bf 49}, 970 (1982);
R. Barbieri, S. Ferrara and C. A. Savoy, Phys. Lett. B {\bf 119}, 343 (1982); 
L. J. Hall, J. D. Lykken and S. Weinberg, Phys. Rev. D {\bf 27}, 2359 (1983); 
P. Nath, R. Arnowitt and A. H. Chamseddine, Nucl. Phys. B {\bf 227}, 121 (1983); 

\bibitem{SUSY-searches1}
 https://twiki.cern.ch/twiki/bin/view/AtlasPublic/SupersymmetryPublicResults
 
 \bibitem{SUSY-searches2}
 https://twiki.cern.ch/twiki/bin/view/CMSPublic/PhysicsResultsSUS
 
 \bibitem{Chatterjee:2011qr}
 A.~Chatterjee and A.~Mazumdar,
  JCAP {\bf 1109}, 009 (2011)
  [arXiv:1103.5758 [hep-ph]].

\bibitem{nG-curvaton} 
  M.~Sasaki, J.~Valiviita and D.~Wands,
  Phys.\ Rev.\ D {\bf 74}, 103003 (2006);
  N.~Bartolo, S.~Matarrese and A.~Riotto,
  Phys.\ Rev.\ D {\bf 69}, 043503 (2004)

\bibitem{Curvaton-enqvist}
 K.~Dimopoulos and D.~H.~Lyth,
  Phys.\ Rev.\ D {\bf 69}, 123509 (2004)
 K.~Enqvist, S.~Kasuya and A.~Mazumdar,
  Phys.\ Rev.\ Lett.\  {\bf 90}, 091302 (2003);
 K.~Enqvist, A.~Jokinen, S.~Kasuya and A.~Mazumdar,
  Phys.\ Rev.\  D {\bf 68}, 103507 (2003);
 K.~Enqvist, S.~Kasuya and A.~Mazumdar,
  Phys.\ Rev.\ Lett.\  {\bf 93}, 061301 (2004);
  R.~Allahverdi, K.~Enqvist, A.~Jokinen and A.~Mazumdar,
  JCAP {\bf 0610}, 007 (2006).
 A.~Mazumdar and A.~Perez-Lorenzana,
  Phys.\ Rev.\ Lett.\  {\bf 92}, 251301 (2004);


\bibitem{Kasuya:2004}
  S.~Kasuya, M.~Kawasaki and F.~Takahashi,
  Phys.\ Lett.\  B {\bf 578}, 259 (2004);
   E.~J.~Chun, K.~Dimopoulos and D.~Lyth,
  Phys.\ Rev.\  D {\bf 70}, 103510 (2004);
  K.~Dimopoulos, G.~Lazarides, D.~Lyth and R.~Ruiz de Austri,
  JHEP {\bf 0305}, 057 (2003);
  K.~Dimopoulos, D.~H.~Lyth, A.~Notari and A.~Riotto,
  JHEP {\bf 0307}, 053 (2003).


\bibitem{Dutta}
S.~Downes, B.~Dutta and K.~Sinha,
  arXiv:1106.2266 [hep-th].
  
\bibitem{Dutta-1}
R.~Allahverdi, S.~Downes and B.~Dutta,
  arXiv:1106.5004 [hep-th].


  \bibitem{B-L}
  R.~Allahverdi, B.~Dutta and A.~Mazumdar,
  Phys.\ Rev.\ Lett.\  {\bf 99}, 261301 (2007)
  [arXiv:0708.3983 [hep-ph]].
  A.~Mazumdar and S.~Morisi,
  arXiv:1201.6189 [hep-ph].

\bibitem{atmos}
R.~N.~Mohapatra, S.~Antusch, K.~S.~Babu, G.~Barenboim, M.~-C.~Chen, A.~de Gouvea, P.~de Holanda and B.~Dutta {\it et al.},
  Rept.\ Prog.\ Phys.\  {\bf 70}, 1757 (2007)
  [hep-ph/0510213].

\bibitem{Maldacena:2002vr} 
  J.~M.~Maldacena,
  JHEP {\bf 0305}, 013 (2003)
  [astro-ph/0210603].

\bibitem{noncanonical}
  E.~Silverstein and D.~Tong,
  Phys.\ Rev.\ D {\bf 70}, 103505 (2004)
  [hep-th/0310221];
    M.~Alishahiha, E.~Silverstein and D.~Tong,
  Phys.\ Rev.\ D {\bf 70}, 123505 (2004)
  [hep-th/0404084];
 X.~Chen, M.~-x.~Huang, S.~Kachru and G.~Shiu,
  JCAP {\bf 0701}, 002 (2007)
  [hep-th/0605045];
    C.~Cheung, P.~Creminelli, A.~L.~Fitzpatrick, J.~Kaplan and L.~Senatore,
  JHEP {\bf 0803}, 014 (2008)
  [arXiv:0709.0293 [hep-th]].
  

\bibitem{Senatore:2009gt} 
  L.~Senatore, K.~M.~Smith and M.~Zaldarriaga,
  JCAP {\bf 1001}, 028 (2010)
  [arXiv:0905.3746 [astro-ph.CO]].

\bibitem{Cai:2010rt} 
  Y.~-F.~Cai and Y.~Wang,
  Phys.\ Rev.\ D {\bf 82}, 123501 (2010)
  [arXiv:1005.0127 [hep-th]].



 
\end{thebibliography}
\end{document}